\documentclass{article}
\usepackage{bm}
\usepackage{bbm}
\usepackage{xcolor}
\usepackage{pifont}
\usepackage{booktabs}
\usepackage{amsmath, amssymb, bigstrut}
\usepackage{amsthm}
\usepackage{mathtools}
\usepackage{apacite}
\usepackage{natbib}
\usepackage{graphicx}
\usepackage{adjustbox}
\usepackage{caption}
\usepackage{lscape}
\usepackage{subcaption}
\usepackage{array, makecell}
\usepackage[linesnumbered,ruled,vlined]{algorithm2e}
\usepackage[utf8]{inputenc}
\usepackage[a4paper, total={6.5in, 9in}]{geometry}

\newsavebox{\mybox}\newsavebox{\mysim}
\newcommand{\distras}[1]{%
  \savebox{\mybox}{\hbox{\kern3pt$\scriptstyle#1$\kern3pt}}%
  \savebox{\mysim}{\hbox{$\sim$}}%
  \mathbin{\overset{#1}{\kern\z@\resizebox{\wd\mybox}{\ht\mysim}{$\sim$}}}%
}
\makeatother

\usepackage[colorlinks=true,citecolor=blue,allcolors=blue]{hyperref} 

\title{\textbf{To Vary or Not To Vary: A Simple Empirical Bayes Factor for Testing Variance Components}}
\author{{Fabio Vieira}$^{1}$ \and Hongwei Zhao$^2$ \and Joris Mulder$^1$ }
 \date{%
     $^1$Department of Methodology and Statistics, Tilburg University\\%
     $^2$Unit of Quantitative Psychology and Individual Differences, KU Leuven\\[2ex]%
     \today
}

\setcitestyle{aysep={}} 
\providecommand{\keywords}[1]
{
  \small	
  \textbf{Keywords:} #1
}

\newcolumntype{R}{@{\extracolsep{0.2cm}}r@{\extracolsep{10pt}}}%

\begin{document}

\RestyleAlgo{ruled}

\maketitle

\begin{abstract}
   Random effects are a flexible addition to statistical models to capture structural heterogeneity in the data, such as spatial dependencies, individual differences, temporal dependencies, or non-linear effects. Testing for the presence (or absence) of random effects is an important but challenging endeavor however, as testing a variance component, which must be nonnegative, is a boundary problem. Various methods exist which have potential shortcomings or limitations. As a flexible alternative, we propose a flexible empirical Bayes factor (EBF) for testing for the presence of random effects. Rather than testing whether a variance component equals zero or not, the proposed EBF tests the equivalent assumption of whether all random effects are zero. The Bayes factor is `empirical' because the distribution of the random effects on the lower level, which serves as a prior, is estimated from the data as it is part of the model. Empirical Bayes factors can be computed using the output from classical (MLE) or Bayesian (MCMC) approaches. Analyses on synthetic data were carried out to assess the general behavior of the criterion. To illustrate the methodology, the EBF is used for testing random effects under various models including logistic crossed mixed effects models, spatial random effects models, dynamic structural equation models, random intercept cross-lagged panel models, and nonlinear regression models.  
\end{abstract}


\section{Introduction}

\par
Structural heterogeneity in the data can be caused by different potential sources, such as, spatial dependencies, temporal dependencies, or individuval differences. Random effects models are the gold standard for capturing the heterogeneity caused by these different sources where the corresponding variance components quantify the degree of the heterogeneity. Random effects models have resulted in numerous insights about the degree of statistical heterogeneity in empirical data. Examples include clinical trials \citep{cleophas2008random}, economics \citep{menegaki2011growth}, finance \citep{alsakka2010random}, social networks \citep{vieira2023bayesian}, meta-analysis \citep{borenstein2010basic}, among others. 

\par
A crucial problem when building random-effect models is to test whether the heterogeneity that is caused by the different sources is sufficient for random effects to be incorporated into the model.
This testing problem serves two purposes. First, understanding the degree of heterogeneity caused by the different sources is important from a substantive perspective. For instance, \cite{hamaker2011model} identified individual differences regarding psychological constructs in longitudinal research using random effects. Second, testing the presence of random effects is of crucial importance when building statistical models for data with complex hierarchical or dependency structures. If a potential source of heterogeneity actually has a negligible impact on the heterogeneity in the data, it is recommended to omit the random effect to avoid statistical overfitting. 

\par
Over the years, many different methods have been proposed to carry out this test \citep{crainiceanu2008likelihood, zhang2008variance, verhagen2013bayesian, roback2021beyond, pauler1999bayes, du2020testing}. A few examples of existing methods include eyeballing, marginal likelihood approaches including Bayes factors, information criteria, likelihood ratio tests, or $k$-fold cross-validation methods. These methods have different potential limitations, such as subjectivity (e.g. eyeballing), computational complexity (e.g. computing marginal likelihoods, cross-validation), problems to control error rates (significance tests), or complex tuning (e.g. choice of the prior in marginal likelihood approaches).

\par
This paper proposes an empirical Bayes factor (EBF) for testing random effects as an alternative to the above approaches with their inherent potential limitations. To circumvent the heavy dependence of the Bayes factor on the prior, which needs to be formulated using external information which is difficult to elicit for random effects variances, an empirical Bayesian approach is considered where the prior is estimated from the data \citep{casella1985introduction,ten1999empirical, liu2019use}. We achieve this using the fact that the distribution of the random effects on the lower level effectively serves as a prior for the random effects, which is estimated from the data when fitting the model. Moreover, to avoid the need to compute marginal likelihoods, which is the main building block of Bayes factors, a Savage-Dickey density ratio is considered using a Gaussian approximation of the posterior which can easily be derived from the output of Bayesian MCMC algorithms and classical algorithms \citep{dickey1971weighted}. The EBF is designed for random effects with any covariance structure, thereby extending the test suggested by \cite{vieira2023bayesian}. 

\par
Due to the extensive literature on testing variance components, we cannot claim that the proposed test is inherently superior to all currently available methodologies (as this depends on how superiority is assessed). The main asset of the proposed test lies in its computational simplicity for this challenging testing problem and its flexibility to test any type of structural heterogeneity in the data under any type of mixed effects model. Thus, the proposed method may very well result in similar conclusions as existing methods for specific testing scenario's and specific data. Its main selling point is notably its simplicity and its general usability.

\par
The remainder of this text is structured as follows: Section \ref{sec:heterog} formally defines the problem of testing heterogeneity; Section \ref{sec:alternatives} discusses the existing alternatives to conduct this test; Section \ref{sec:EBF} introduces the Empirical Bayes Factor; Section \ref{sec:simulation} presents a synthetic data study showing the behavior of the EBF; \ref{sec:application} shows how the EBF can be applied to different settings and to models in different classes; and finally Section \ref{sec:discussion} ends up with concluding remarks and a discussion. 

\section{Testing heterogeneity}\label{sec:heterog}

Let us consider a general formulation of a mixed effect model,
\begin{eqnarray}
    \label{eq:model}
        \bm{y} &\sim& p(\bm{X}, \bm{\phi}, \bm{\theta}_1,\ldots,\bm{\theta}_K)\\
\nonumber        \bm{\theta}_k &\sim& \mathcal{N}(\bm{0}, \bm{\Psi}_k(\bm{\gamma}_k)),\text{ for }k=1,\ldots,K,
\end{eqnarray}
\noindent
where $\bm{y}$ is the response vector and $\bm{X}$ is the data (design) matrix with relevant predictors and covariates, $\bm{\phi}$ is a vector of nuisance parameters, and $\bm{\theta}_k$ is a vector of length $J_k$ containing the group (or cluster) specific random effects originating from the $k$-th source of heterogeneity. Its distribution is parameterized by a structured covariance matrix $\bm{\Psi}_k(\bm{\gamma}_k)$ with $L_k$ cluster-specific variance components which are contained in the vector $\bm{\gamma}_k$, for $k=1,\ldots,K$. Our goal is to test the (non)variability of the specific random effects. In other words, for a given variance component $\gamma_{kl}$, we want to evaluate which of the two models fits the data best:
\begin{eqnarray}
    \label{eq:testingproblem}
\text{M}_{0}&:& \gamma_{kl} = 0\\
\nonumber\text{M}_{1}&:& \gamma_{kl} > 0,
\end{eqnarray}
where all other model parameters are nuisance parameters.
For each variance component, such a model selection problem can be formulated. Moreover, it would also be possible to test multiple random effects jointly (although we mainly focus on testing the separate random effects as this is more informative).

\par
Practically any statistical model is a special case of the general formulation of the model in \eqref{eq:model}, such as generalized linear mixed effects models, dynamic structural equation models, or spatial random effects models. Section \ref{sec:application} describes various examples.
Here we give two simple examples to illustrate the parameterization and cluster indexing.
First, we consider a model with two crossed random intercepts, which, thus, contains two potential sources of heterogeneity, and which can be written as
\begin{equation}
\begin{gathered}
    y_{ijk} \sim \mathcal{N}(\mu + \theta_{1j} + \theta_{2k}, \sigma^2)\\
    \bm{\theta}_1 \sim \mathcal{N}(0, \tau^2_{\theta1} \bm{I}_{J_1})\\
    \bm{\theta}_2 \sim \mathcal{N}(0, \tau^2_{\theta2} \bm{I}_{J_2}).\\
\end{gathered}
\end{equation}
\noindent
Both cluster dimensions contain one variance component, i.e., $\gamma_{11}=\tau^2_{\theta1}$ and $\gamma_{21}=\tau^2_{\theta2}$ with $L_1=L_2=1$. The nuisance parameters are then the global mean $\mu$ and the within cluster variance $\sigma^2$. Two model selection problems can be formulated on the two variance components: (i) $\text{M}_0: \tau^2_{\theta1}=0$ versus $\text{M}_1:\tau^2_{\theta1}>0$ (where $\tau^2_{\theta2}$ is a nuisance parameter that is unconstrained) and (ii) $\text{M}_0: \tau^2_{\theta2}=0$ versus $\text{M}_1:\tau^2_{\theta2}>0$ (where $\tau^2_{\theta1}$ is a nuisance parameter that is unconstrained).

\par
In a random intercept random slope model, there is one source of heterogeneity, $K=1$, with two variance components, $\bm\gamma_1=(\tau_{\theta1},\tau_{\theta2})'$ with $L_1=2$.  A random intercept random slope model can be formulated as a special case of Equation \eqref{eq:model} as follows
\begin{eqnarray*}
    y_{ij} &\sim& \mathcal{N}(\beta_1 + \theta_{1j} + (\beta_2+ \theta_{2j} )\bm{x}_{ij}, \sigma^2)\\
    \begin{bmatrix}
        \theta_{1j} \\
        \theta_{2j}
    \end{bmatrix} &\sim&
    \mathcal{N}\Bigg(\bm{0},     \begin{bmatrix}
        \tau^2_{\theta1} \ \ \rho\tau_{\theta1}\tau_{\theta2}\\
        \rho\tau_{\theta1}\tau_{\theta2} \ \ \tau^2_{\theta2}
    \end{bmatrix}
    \Bigg),
\end{eqnarray*}
\noindent
The nuisance parameters are the correlation $\rho$ represents the covariance between the random slopes $\theta_{2j}$ and random intercepts $\theta_{1j}$, the fixed intercept $\beta_1$ and fixed slope $\beta_2$, and the within cluster variance $\sigma^2$. The interest would be in testing for the presence of random intercepts, i.e., $\text{M}_0: \tau^2_{\theta1}=0$ versus $\text{M}_1:\tau^2_{\theta1}>0$ (where the random slopes variance $\tau^2_{\theta2}$ would serve as nuisance parameters) and the presence of random slopes, i.e., $\text{M}_0: \tau^2_{\theta2}=0$ versus $\text{M}_1:\tau^2_{\theta2}>0$ (where the random slopes variance $\tau^2_{\theta1}$ would serve as nuisance parameters).

\par

\par

\par

\par

\section{Existing methods}\label{sec:alternatives}

In this section, we give a brief overview of the most commonly used model selection criteria that have been proposed in the literature. We also discuss the potential limitations of these methods, in particular for the current testing problem on the variability of random effects. To simplify the notation when deriving these different criteria, we shall consider a model with a single random effect, so that we can omit the cluster indices, and the model selection problem in \eqref{eq:testingproblem} can be written as $\text{M}_{0}: \gamma = 0$ versus $\text{M}_{1}: \gamma > 0$.

\subsection{Eyeballing}

Eyeballing is the practice of drawing conclusions by simply looking at the estimates of the key parameter. It is a common practice in exploratory data analysis where visual inspections can be a useful starting point \citep{tukey1977exploratory, hartwig1979exploratory}. However, eyeballing can be highly misleading when used to conduct statistical inference. \cite{ioannidis2008interpretation} refers to eyeballing as a pseudo-test that only reflects the biases and speculations of the analyst. Moreover, the lack of a formal statistical framework might lead to deceptive confidence in inferences that are, at best, pseudo-scientific. 

\par
Although eyeballing is generally discouraged as a formal tool to conduct inferences \citep{sharpe2015chi}, some researchers, such as \cite{ludbrook2011there}, still vouch for it. Sometimes even going against traditional statistical tests, even though there is evidence that experienced researchers fail in the eyeballing test \citep{terrin2005empirical}. Eyeballing was for instance used by \cite{mcneish2021measurement} to assess the presence of heterogeneity in a dynamic structural equation model with crossed random effects.

\par
A slightly more sophisticated approach would be to eyeball interval estimates (either classical confidence bounds or Bayesian credibility bounds). However, eyeballing the interval estimates when testing variance components would not give any insights as variances are positive by definition, and thus zero will never lie in the interval. 

\subsection{Marginal modeling approaches}

A marginal modeling approach can be adopted by integrating out the random effects of interest

\begin{equation}
    p(\bm{y} | \textbf{X}, \bm{\phi}, \bm{\gamma}) = \int p(\bm{y} | \textbf{X}, \bm{\phi}, \bm{\theta}) \mathcal{N}(\bm{\theta} | \bm{0}, \bm{\Psi}(\bm{\gamma})) d \bm{\theta}. 
\end{equation}

\noindent
In this framework, the random-effect variance components $\bm\gamma$ can becomes covariances which can be negative. Thus, checking whether zero lies in the interval would be possible. For example, for simple random intercept designs, the computation of the posterior probability that the variance is positive is relatively easy using the proportion of draws satisfying this constraint. Note that a marginal modeling approach is also considered when fitting structural equation models \citep[SEMs,][]{rosseel2012lavaan} by integrating out the latent variables.

\par
 To our knowledge, a marginal modeling approach for testing random effects variances has only been proposed for very specific models, such as random intercept models \citep{mulder2013bayesian, fox2017bayes}, item response models \citep{fox2017bayes}, and simple random intercept and random slope models \citep{nielsen2021small}. In a Bayesian approach the complexity of this approach lies in the fact that sampling the variance components (which are covariance after integrating out the random effects) can be difficult as complex restrictions need to hold in order for the structured covariance matrix of the data to be positive definite under the integrated model.
 

\subsection{Information criteria}

Information criteria (IC) are statistical measures that balance model fit and model complexity to assess the adequacy of the model. Thus, when two models fit the data approximately equal the simplest model would be preferred as it gives a simpler explanation of the data generating mechanism. Typically model complexity is quantified via the number of free parameters under a specific model. The model with the lowest outcome of an IC is generally preferred. The IC of model M can be written in a general form as follows:
\begin{equation}
\label{eq:IC}
    \text{IC}(\text{M}) = -2 \log (p(\bm{y} | \bm{X}, \tilde{\bm{\phi}}, \tilde{\bm{\theta}}, \tilde{\bm{\gamma}})) + \lambda,
\end{equation}
\noindent
where $\tilde{\bm{\phi}},\ \tilde{\bm{\theta}}$, and $\tilde{\bm{\gamma}}$ are model estimates (e.g. maximum likelihood estimates). In this expression, the first term captures the model fit and the second part, $\lambda$, is a penalty for model complexity that differs among the criteria \citep{hamaker2011model}.

\par
The Akaike information criterion (AIC) is probably the most famous information criterion. Introduced by \cite{akaike1998information}, it has an informational theoretical background. The goal of the AIC is, under certain conditions, to seek among the models proposed the one that minimizes the Kullback-Leibler distance between the true data-generating distribution and the estimated model \citep{anderson2002avoiding, wood2017generalized}. The AIC formulas, for the selection of best model between $\text{M}_{0}: \bm{\Psi}(\bm{\gamma}) = \bm{0}$ or $\text{M}_{1}: \bm{\Psi}(\bm{\gamma}) > \bm{0}$, are given by
\begin{equation}
\label{eq:AIC}
\begin{gathered}
    \text{AIC}(\text{M}_0) = -2 \text{log}(p(\bm{y} | \bm{X}, \hat{\bm{\phi}}_0)) + 2 q_0\\
    \text{AIC}(\text{M}_1) = -2 \text{log}(p(\bm{y} | \bm{X}, \hat{\bm{\phi}}_1, \hat{\bm{\theta}}_1, \hat{\bm{\gamma}}_1)) + 2 q_1,
\end{gathered}
\end{equation}
\noindent
where $p(\bm{y} | \bm{X}, \hat{\bm{\phi}}_0)$ and $p(\bm{y} | \bm{X}, \hat{\bm{\phi}}_1, \hat{\bm{\theta}}_1, \hat{\bm{\gamma}}_1)$ are the respective maximized likelihoods evaluated at the maximum likelihoods estimates (MLEs) under the respective models. In this criterion, the penalty $\lambda$ from Equation \eqref{eq:IC}, is equal to $2q_t$, where $q_t$ is the number of parameters in model M$_t$, for $t=0$ or 1.

\par
The Bayesian information criterion (BIC) is based on the marginal likelihood using a Laplace approximation and an (implicit) unit information prior for the free parameters \citep{schwarz1978estimating, raftery1995bayesian}. To compute the BICs for the two models of interest, the following formulas are used
\begin{equation}
\label{eq:BIC}
\begin{gathered}
    \text{BIC}(\text{M}_0) = -2 \text{log}(p(\bm{y} | \bm{X}, \hat{\bm{\phi}}_0)) + \text{log}(N) q_0\\
    \text{BIC}(\text{M}_1) = -2 \text{log}(p(\bm{y} | \bm{X}, \hat{\bm{\phi}}_1, \hat{\bm{\theta}}_1, \hat{\bm{\gamma}}_1)) + \text{log}(N) q_1.
\end{gathered}
\end{equation}
\noindent
In this case, the penalty is expressed as $\lambda = \log(N) q_t$, and $N$ is the number of observations in the data set and $q_t$ is, again, the number of parameters in model M$_t$.

\par
A third information criterion, which is commonly in the Bayesian literature, is the deviance information criterion \citep{spiegelhalter2002bayesian}. If one uses an approximately normal likelihood with vague priors, then the DIC and AIC are equivalent \citep{hamaker2011model}. However, as opposed to the AIC, which uses information theory, the DIC has a decision theoretical background where the goal is to find the model that minimizes the expected posterior loss. The formulas to compute the DICs are given by
\begin{equation}
\begin{gathered}
    \text{DIC}(\text{M}_0) = -2\ \log(p(\bm{y} | \bm{X}, \bar{\bm{\phi}}_0)) + 2\ \text{P}_{\text{DIC}(\text{M}_0)}\\
    \text{DIC}(\text{M}_1) = -2\ \log(p(\bm{y} | \bm{X}, \bar{\bm{\phi}}_1, \bar{\bm{\theta}}_1, \bar{\bm{\gamma}}_1)) + 2\ \text{P}_{\text{DIC}(\text{M}_1)},
\end{gathered}
\end{equation}
\noindent
where $\bar{\bm{\phi}}_t, \bar{\bm{\theta}}_t, \bar{\bm{\gamma}}_t$ are posterior means under model M$_t$, for $t=0$ or 1, and the penalties reflect the effective number of parameters which are given by
\begin{equation}
\begin{gathered}
\label{eq:Pdic}
    P_{\text{DIC}(\text{M}_0)} = 2 \Big(\log(p(\bm{y} | \bm{X}, \bar{\bm{\phi}})) - \frac{1}{S} \sum_{s = 1}^{S} \log(p(\bm{y} | \bm{X}, \bm{\phi}^s)) \Big)\\
    P_{\text{DIC}(\text{M}_1)} = 2 \Big(\log(p(\bm{y} | \bm{X}, \bar{\bm{\phi}}, \bar{\bm{\theta}}, \bar{\bm{\gamma}})) - \frac{1}{S} \sum_{s = 1}^{S} \log(p(\bm{y} | \bm{X}, \{\bm{\phi}, \bm{\theta}, \bm{\gamma}\}^s) \Big),
\end{gathered}
\end{equation}
\noindent
where the superscript $s$ denotes the $s^{th}$ draw in a Markov chain Monte Carlo sample.

\par
Models often contain multiple random effects to capture different sources of potential heterogeneity. Thus, to test all these different random effects, ICs have to be computed for all possible combinations of included/excluded random effects. This can be computationally demanding in the case many potential sources of heterogeneity are present. For example, in the case of $P$ different random effects, $2^P$ models would need to be fit (in a exploratory setup) to assess which random effects model is most adequate for the data at hand. Another potential issue, specifically associated with the AIC and the BIC, is that both require counting the number of parameters in the model. It is unclear whether the random effects should be treated as free parameters as their distribution is specified on a lower level. For the BIC, it may also not be clear whether the number of clusters, the total sample size, or a measure of the effective sample size should be used for the sample size $N$. Moreover, the BIC and the AIC require the availability of maximum likelihood estimates of the model parameters which may not be available, especially in more complex models involving higher-order structures with possible crossed random effects and latent variables. This fact limits the applicability of the AIC and the BIC for more complex models such as dynamic structural equation models \citep{asparouhov2018dynamic}.

\par
The DIC overcomes the issue how to compute the number of free parameters as this is estimated from the fitted model. However, an important limitation remains: one still needs to fit many models with different combinations of random effects present which can be computationally demanding. Moreover, we can see in equation \ref{eq:Pdic} that if $\log(p(\bm{y} | \bm{X}, \bar{\bm{\phi}}, \bar{\bm{\theta}}, \bar{\bm{\gamma}})) < \frac{1}{S} \sum_{s = 1}^{S} \log(p(\bm{y} | \bm{X}, \bm{\phi}, \bm{\theta}, \bm{\gamma})^s)$, then the effective number of parameters becomes negative, i.e., $P_{\text{DIC}} < 0$, which contradicts with its interpretation. To ensure that estimated number of model parameters is generally greater than zero, \cite{gelman2014understanding} proposed an alternative to the $P_{\text{DIC}}$ based on the variance of the logarithm of the posterior distribution, which would force the effective number of parameters to be always positive. As stated by these authors, this measure can be less numerically stable than the regular $P_{\text{DIC}}$.

\subsection{Likelihood ratio tests}

The likelihood ratio test (LRT) is a statistical significance test commonly used to compare nested models. The test statistic of model M$_1$ relative to M$_0$ can be written as follows:
\begin{equation}
\label{eq:LRT}
    \text{LRT}(\text{M}_1, \text{M}_0) = 2 (\text{log} (p(\bm{y} | \bm{X}, \hat{\bm{\phi}}, \hat{\bm{\theta}}, \hat{\bm{\gamma}}) - \text{log} (p(\bm{y} | \bm{X}, \hat{\bm{\phi}}))),
\end{equation}
\noindent
where $\hat{\bm{\phi}}, \hat{\bm{\theta}}$, and $\hat{\bm{\gamma}}$ are maximum likelihood estimates. \cite{stram1994variance} have shown that the asymptotic distribution of this statistic is an equal-weighted mixture of chi-squared distributions. This asymptotic distribution is used to determine a cut-off value when to reject the simpler model given a prespecified significance level which controls the type I error to incorrectly reject the simpler model. In statistical practice however `asymptopia' (i.e. the fairy tale land where data is unlimited and estimators are always consistent) is often not reached. Thus, the use of the asymptotic distribution can result in inaccurate type I errors \citep{leamer2010tantalus, abbott2022far}. Appendix \ref{app:chi_squared} of this manuscript also provides an example which is discussed later in Section 6.4. Hence, the reported type I error rates, which is the main focus of the test, may be inaccurate. Therefore, LRTs should be used with extreme care.

\par
For the current testing problem of random effects, it has been shown that the sampling 
distribution of the LRT statistic indeed often does not follow the presumed mixture of Chi-squared distributions (for a discussion see \cite{crainiceanu2008likelihood}, and \cite{zhang2008variance}). Therefore, in practice, this test is usually carried out via parametric bootstrapping. This  can be very computationally intensive as bootstrap-based tests normally require a large number of bootstrap samples and each model needs to be fitted to each sample resulting in a huge number of fitted models \citep[see Chapter 11]{roback2021beyond}.

\subsection{K-fold cross-validation}

\par
Cross-validation (CV) is a popular method used for quantifying a model's out-of-sample prediction error. The technique consists of separating the data into $K$ sets of training and test data. The training set is used to fit the model and the test set is used to evaluate its predictive performance \citep{berrar2019cross}. This procedure is then repeated until all observations have been used for testing. In the general case, the data set of $N$ observations is divided into $K$ partitions, where $K-1$ are used for training, the average prediction error is given by 
\begin{equation}
\label{eq:cv}
    \text{CV}(\text{M}) = \frac{1}{N} \sum_{k = 1}^{K} \sum_{i \in k} \epsilon_{i}^{(k)}
\end{equation}
\noindent
where $\epsilon_{i}^{(k)}$ is the error metric of interest, for observation $i = 1, \dots, N$ in partition $k = 1. \dots, K$. For instance, it can be the squared error $\epsilon_{i}^{(k)} = {(\hat{\bm{y}}_{i}^{(k)} - \bm{y}_{i}^{(k)})}^2$, or the absolute error $\epsilon_{i}^{(k)} = |\hat{\bm{y}}_{i}^{(k)} - \bm{y}_{i}^{(k)}|$, where $\hat{\bm{y}}_{i}^{(k)}$ is the model prediction for observation $i$ and  $\bm{y}_{i}^{(k)}$ is the true value of observation $i$ on partition $k$. Then, the model with the lowest $\text{CV}(M)$ is the preferred model.

\par
An attractive property of cross-validation is its interpretation as measure of predictive performance and its flexibility (without relying on asymptotic distributions). Thus, it can be a useful tool, especially in machine learning research where the focus is mostly on predictive performance \citep{schaffer1993selecting}. However, the main issue with using $K$-fold cross-validation is how to select $K$  \citep{anguita2012k}. Popular choices are $K = 3,\ 5,\ \text{and}\ 10$, however, these choices are quite arbitrary. To circumvent the problem of choosing $K$, one can simply make $K = N$, yielding a special case of $K$-fold cross-validation called leave-one-out cross-validation \citep{gelman2014understanding, vehtari2012survey, vehtari2020loo}. There have been studies of this particular case showing that it is not consistent as the probability of choosing the true model does not go to one as $N \rightarrow \infty$ \citep[for instance]{shao1993linear}. \cite{gronau2019limitations} have studied the particular case of leave-one-out cross-validation in a specific setup when comparing a simpler model against a more complex one. They concluded that cross-validation fails to strongly endorse the simpler model even when the data aligns perfectly with it and the sample size grows to infinity. 
An alternative could be to try different values of $K$ among the popular choices until we find the best one. But, this could be computationally demanding and time-consuming.

\subsection{Bayes factors}\label{subsec:BF}

The Bayes factor (BF) is a statistical criterion that quantifies the relative evidence provided by the data between two models \citep{jeffreys1961theory}. For the current setup, the Bayes factor for model $\text{M}_{0}$ against model $\text{M}_{1}$ is defined as the ratio of their respective marginal likelihoods:
\begin{equation}
\label{eq:BF}
    \text{BF}_{01} = \frac{m_0(\bm{y} | \bm{X})}{m_1(\bm{y} | \bm{X})},
\end{equation}

\noindent
where
$$m_0(\bm{y} | \bm{X}) = \int p(\bm{y} | \bm{X}, \bm{\phi})\ p(\bm{\phi}) d \bm{\phi},\\$$
$$m_1(\bm{y} | \bm{X}, \bm{\phi}, \bm{\theta}, \bm{\gamma}) = \int \int \int p(\bm{y} | \bm{X}, \bm{\phi},\bm\theta)\ \mathcal{N}(\bm{\theta} | \bm{0}, \bm{\Psi}(\gamma))\ p(\bm{\phi})\ p(\bm{\gamma}) d \bm{\phi} d \bm{\theta} d \bm{\gamma},$$
are the marginal likelihood for models $\text{M}_0$ and $\text{M}_1$, respectively, reflecting the probability of the data under the two respective models.

\par
When computing Bayes factors there are two important challenges that need to be addressed. Firstly, the specification of proper priors (improper priors cannot be used, e.g., \cite{o1995fractional}) that quantify which values of the unknown model parameters are most likely before observing the data. This step is important as marginal likelihoods and Bayes factors are very sensitive to the exact choice of the prior (much more than the posterior in Bayesian estimation), especially to the prior of the key parameters that are being tested, which are the variance parameters in $\bm{\Psi}(\bm{\gamma})$ \citep{sinharay2002sensitivity, liu2008bayes, vanpaemel2010prior}. Choosing this prior for a given context using external knowledge is very challenging as the interpretation of the values depend on many different factors such as the scale of the data and the nuisance parameters that are present. For instance, \cite{mulder2019bayes} reparameterized a random-effect model using intraclass correlations, which are bounded, and then proper uniform priors were used as a default choice. It is yet unclear how this can be extended to a general setting, as we consider here.

\par
Secondly, once proper priors are specified, the challenge is to compute the marginal likelihoods. For random-effect models, which usually have complex likelihoods with a large number of parameters, this task can be very challenging. Even though various approximations to the marginal likelihood have been proposed (e.g. \cite{gelman1998simulating, geweke1999using, friel2008marginal}), marginal likelihood computation remains a time-consuming task that hinders the method's usefulness for more complex models. Thus, this computational burden would create another layer of complexity that would prevent practitioners from using Bayes factors for testing the inclusion of random effects.

\section{The empirical Bayes factor}\label{sec:EBF}

In this section, we present the empirical Bayes factor as a flexible alternative to test, select, and build random effects models. Again, we use the simplified notation of testing a single variance component $\gamma$ with random effects $\bm\theta$, and thus omitting the index $k$. Two properties of mixed effects models are used to arrive at the empirical Bayes factor. First, under the restricted model $M_0$, the random effects covariance matrix is not positive definite, while under the larger unconstrained model $M_1$, the random effects covariance matrix is positive definite. Thus, the models can equivalently be written as a restricted model where the random effects are zero and an unrestricted where the random effects are unequal to zero, i.e.,
\begin{eqnarray}
\label{M0M1alt}
\text{M}_0&:& \bm{\theta}=\bm{0}\\
\nonumber    \text{M}_1&:& \bm{\theta}\in\mathbb{R}^J.
\end{eqnarray}
Second, we use the fact that the Gaussian distribution of the random effects on the lower level under $M_1$ has the same role as a prior in a Bayesian model. For this reason, classical estimates of mixed effects models are also commonly referred to empirical Bayesian approaches \citep[e.g.,][]{casella1985introduction}. As the random effects distribution is part of the model, it is fitted from the observed data, and thus we can write
\begin{equation}
\bm{\theta} | M_1 \sim \mathcal{N}(\bm{0}, \bm{\Psi}(\tilde{{\gamma}})),
\label{estprior}
\end{equation}
where an estimate is used for the variance components, denoted by $\tilde{\bm{\gamma}}$, which can be a Bayesian or classical estimate. Additionally, when fitting a mixed effects model (either using classical or Bayesian algorithms), the estimated random effects and its uncertainty is based on a combination the information from the lower (prior) level \eqref{estprior} and the information on the first (likelihood) level in \eqref{eq:model}. In a Bayesian perspective, the resulting distribution is referred to as the posterior of the random effects $\bm\theta$ under $M_1$. The marginal posterior can be approximated using a multivariate Gaussian distribution, i.e.,
\begin{equation}
\bm{\theta} | M_1,\bm{y} \sim \mathcal{N}(\tilde{\bm{\theta}}, \tilde{\bm{\Omega}}_{\theta}),
\end{equation}
where $\tilde{\bm{\theta}}$ denotes the estimated random effects and $\tilde{\bm{\Omega}}_{\theta}$ denotes its uncertainty covariance matrix.

Using the prior and posterior for the key parameters under the larger unconstrained model $M_1$, a simple methodology to compute a Bayes factor between a restricted model where all parameters are zero against an unrestricted alternative is the Savage-Dickey density ratio \citep{dickey1971weighted,wagenmakers2010bayesian}. The Savage-Dickey density ratio is defined as the ratio of the posterior and prior density of the key parameters at the null value under the unrestricted model M$_1$. This method has a key computational advantage that the direct computation of marginal likelihoods under the respective models can be avoided. Only the fitted unconstrained model is required. Because the prior and posterior follow multivariate Gaussian distributions, the Savage-Dickey density ratio has a simple analytical expression. The Bayes factor of model $M_0$ against $M_1$ can then simply be expressed as 
\begin{eqnarray}
\nonumber
\text{EBF}_{01} &=& \frac{\pi({\bm\theta}= \bm 0 | \bm{y},M_1)}{\pi({\bm\theta}=\bm 0|M_1)}\\
\nonumber &=&\frac{\mathcal{N}(\bm{0};\tilde{\bm{\theta}}, \tilde{\bm{\Omega}}_{\theta})}{\mathcal{N}(\bm{0};\bm{0}, \bm{\Psi}(\tilde{{\gamma}}))}\\
\label{eq:SD_bf} &=& |\bm{\Psi}(\tilde{{\gamma}})|^{1/2} |\tilde{\bm{\Omega}}_{\theta}|^{-1/2}
\exp\{-\tfrac{1}{2}\tilde{\bm{\theta}}'\tilde{\bm{\Omega}}_{\theta}^{-1}\tilde{\bm{\theta}}
\}.
\end{eqnarray}
As the prior is based the observed data, similar as empirical Bayesian approaches, we shall refer to \eqref{eq:SD_bf} as the empirical Bayes factor for testing variance components. To our knowledge, empirical Bayes factors have not yet been proposed for testing variance components in mixed effects models although empirical Bayes factors have been proposed for testing coefficients in regular regression models using prior hyperparameters that are estimated from the data \citep{george2000calibration}. Moreover, empirical Bayesian approaches are common in regularization problems \citep[e.g.,][]{van2019shrinkage}.

To obtain the estimates for the variance components, $\tilde{\gamma}$, the random effects, $\tilde{\bm{\theta}}$, and its error covariance matrix, $\tilde{\bm{\Omega}}_{\theta}$, either classical or Bayesian procedures can be used. The R package \texttt{lme4} \citep{bates2015package} can be used to obtain classical estimates for generalized linear mixed effects models. To our knowledge, the error covariances of the random effects are not provided by the output of \texttt{lme4} functions however. For this reason, a diagonal error covariance matrix can be used having the squared standard errors in the diagonals. For other types of mixed effects models, other R packages can be used. To obtain Bayesian estimates, the R package \texttt{brms} \citep{burkner2017brms} could be used for fitting a large variety types of mixed effects models. Many more packages are available as well of course for fitting specific mixed effects models, as well as very general packages such as \texttt{rstan} \citep{guo2020package} or \texttt{rjags} \citep{plummer2012jags}. Regarding the choice of the priors of the variance components $\bm\gamma$, we recommend using flat noninformative priors to avoid prior shrinkage in a certain (arbitrary) direction. For the variance of a random intercept $\tau^2$, this improper prior can be written as $\pi(\tau^2)\propto 1$, which was recommended by \cite{chung2013variance} as a default choice in estimation problems\footnote{It can be shown that the improper gamma(2,0) prior for the random effects standard deviation $\tau$, as recommended by \cite{chung2013variance}, is equivalent to a flat prior on the variance $\tau^2$.}. Note here that because the variance components serve as nuisance parameters in the above model selection problem \eqref{M0M1alt}, improper priors can be used for computing the empirical Bayes factor.

Finally, we highlight some properties of the proposed empirical Bayes factor in relation to the properties of the methods that were discussed in Section \ref{sec:alternatives}. First, the proposed Bayes factor also relies on a Gaussian approximation similar as the BIC. The BIC however uses a Gaussian approximation for the entire likelihood \citep{raftery1995bayesian}, while the proposed EBF only uses a Gaussian approximation of the posterior of the parameters that are tested. Thus, the proposed empirical Bayes factor relies less heavily on the Gaussian approximation than the BIC. Second, unlike regular Bayes factors which are highly sensitive to the choice of the prior of the tested parameters, the proposed EBF avoids this as the `prior' is estimated from the data when fitting the full model, and thereby, abiding the principle of empirical Bayesian approaches. Third, in order to obtain all EBFs of the full model against all restricted models where specific random effects are fixed, only the fitted full model is required and the restricted models do not have to be fitted unlike many of the other approaches that were discussed in Section \ref{sec:alternatives}. Thus, the EBF is extremely simple and fast to compute for all random effects model that are present in the model. Before illustrating this in various applications in Section \ref{sec:application}, the general behavior of the EBF is assessed in a synthetic data study.

\section{Synthetic data study}\label{sec:simulation}
The general behavior of the EBF is assessed under a controlled setting using a specific mixed effects model. The goal of this synthetic study was not to check classical error rates under different `true' data generating models because these error rates would differ depending on many (arbitrary) choices regarding the data (e.g. sample sizes, possible covariates, measurement levels of the variables, etc.), regarding the complexity of the true model (the number of parameters, the dependency structures across parameters, the number of levels, the possible presence of crossed effects and/or latent variables, etc.), and the regarding the arbitrary cut-off that would be used. For this reason, the usefulness of classical error rates under a very specific settings is of very limited use in general statistical practice, especially because of the general applicability of the EBF under virtually any statistical model that contains random effects (which are becoming increasingly complex due to advanced statistical software). Obtaining accurate error rates in more realistic but highly complex models is also not computationally feasible (for example, fitting the dynamic structural equation model to empirical data in Section 6.3 took about 3 weeks). In addition, Bayes factors are statistically consistent under very general conditions implying that the error rates decrease as the sample size grows. Therefore, the goal of this synthetic study is mainly to see how the evidence for the presence of random effects depends on different (controlled) characteristics of the data, such as the number of clusters, the cluster sizes, the degree of heterogeneity, and the estimates that are used to compute the EBFs (either classical or Bayesian).

To achieve this, the following crossed random effects model with two sources of heterogeneity, each with a random intercept and a random slopes, was considered
\begin{eqnarray}
\label{simmod}y_{ijk} &=& \alpha_1 + \theta_{11j}+\theta_{21k}+(\alpha_2+\theta_{12j}+\theta_{22k})x_{ijk}+\epsilon_{ijk}\\
\nonumber(\theta_{11j},\theta_{12j}) &\sim & 
\mathcal{N}\Bigg(\bm{0},     \begin{bmatrix}
        \tau^2_{11} \ \ \rho_{1}\tau_{11}\tau_{12}\\
        \rho_{1}\tau_{11}\tau_{12} \ \ \tau^2_{12}
    \end{bmatrix}
    \Bigg)\\
\nonumber(\theta_{21k},\theta_{22k}) &\sim &
\mathcal{N}\Bigg(\bm{0},     \begin{bmatrix}
        \tau^2_{21} \ \ \rho_{2}\tau_{21}\tau_{22}\\
        \rho_{2}\tau_{21}\tau_{22} \ \ \tau^2_{22}
    \end{bmatrix}
    \Bigg)\\
\nonumber\epsilon_{ijk} &\sim& \mathcal{N}(0,\sigma^2),
\end{eqnarray}
for $i=1,\ldots,n$, $k=1,\ldots,K$, and $j=1,\ldots,J$. Data were generated by first generating random errors using $\sigma^2=1$, which are then scaled to have an exact sample mean of 0 and an exact sample variance of 1. The same was done for the random effects which then had sample means of 0, and sample standard deviations of $\hat{\tau}_{21}=0.5$, $\hat{\tau}_{22}=0.01$, $\hat{\tau}_{12}=0.5$. Moreover, the sample standard of the first random intercept, $\hat{\tau}_{11}$, was varied from 0 to 1. The random effects correlations were 0 in the data (although other values did not qualitatively affect the results). The covariates, $x_{ijk}$, were generated to have a mean of 0 and a variance of 1. Finally, the number of clusters for the first dimension were varied between $J=10,$ 30, or 100, and the cluster sizes varied between $n=10$, 30, or 100. To obtain Bayesian estimates, a Bayesian fit of model \eqref{simmod} was obtained using \texttt{rstan} \citep{rstan2024} with uniform prior for the variance components. To obtain classical estimates, the R package \texttt{lme4} \citep{bates2015package}) was used to fit the model where the second random slope was omitted because it resulted in singular fits (due to its negligible contribution in the data, i.e., $\hat{\tau}_{22}=0.01$), and the smallest sample standard deviation of $\hat{\tau}_{11}$ was set to 0.04 (also to avoid singular fits) when using \texttt{lme4}. Therefore, using the classical estimates, we cannot compute the EBFs to test the second random slope and we cannot compute the EBFs in the extreme setting when the first random intercept $\hat{\tau}_{11}$ was exactly 0 (i.e., where the effects are fixed).

\begin{figure}
    \centering
    \includegraphics[width = 16.5cm]{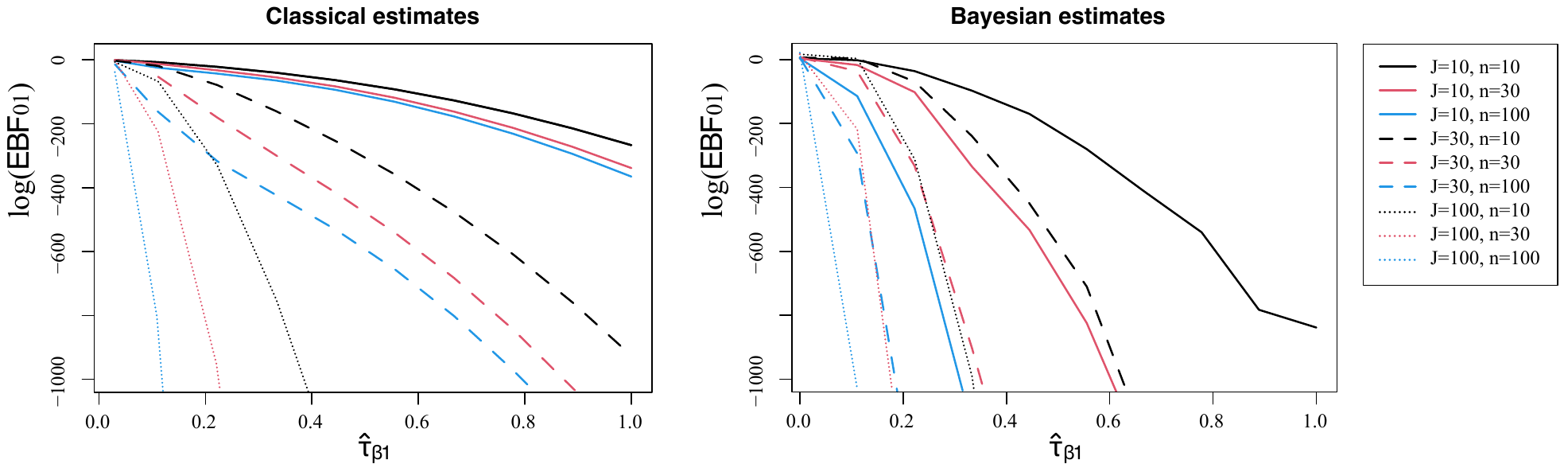}\\
    \caption{Logarithm of the EBF for testing $M_0:\tau_{11}^2=0$ versus $M_1:\tau_{11}^2>0$ for different numbers of clusters $J$, different cluster sizes $n$, different sample standard deviations of the random effects in the data $\hat{\tau}_{11}$, and using classical MLEs (left panel) and Bayesian posterior means (right panel).}
    \label{fig:simulation_bayes}
\end{figure}

Figure \ref{fig:simulation_bayes} shows the logarithm of the EBF for testing the first random intercept, i.e., $M_0:\tau^2_{11}=0$ against $M_1:\tau^2_{11}>0$, for different numbers of clusters $J$, different cluster sizes $n$, different sample standard deviations of the random effect in the data, $\hat{\tau}_{11}$, and using classical MLEs (left panel) and Bayesian posterior means (right panel). The figures show the anticipated decreasing trend of the EBF when the sample standard deviation increases from 0 (no heterogeneity caused by the first random intercept) to 1 (high heterogeneity caused by the first random intercept). Moreover, we see that in general more clusters and larger clusters result in more evidence for a random effect when the data are also heterogenous. In the extreme case that the sample standard deviation $\hat{\tau}_{11}$ is very close to zero (exactly 0 is not possible due to singular fits), the classical estimates result in a logarithm of the EBF around 0 implying neither evidence for or against a random effect. When using Bayesian estimates on the other hand, we do obtain some evidence in favor of M$_0$ in the extreme case that $\hat{\tau}_{\beta1}=0$. Thus the classical estimates result in more liberal EBFs towards $M_1$.

To better see the differences in this extreme setting, Table \ref{tab:simulation} shows the EBFs for testing the first random intercepts for the smallest standard deviation of the first random intercept in the data. The difference in EBFs using the classical and Bayesian estimates can be explained by the skewness of the likelihood (and thus of the posterior). Consequently, the posterior means will be larger than MLEs. Furthermore, the evidence for a null model using a Bayes factor will be larger when the prior standard deviation is larger, which is the case when the posterior mean is used as estimate of the random effects variance on the lower (`prior') level of the model. For this reason, it would be recommendable to interpret the EBFs slightly more conservatively when they are computed using classical estimates. Finally, Table \ref{tab:simulation} also shows the EBFs when testing the other random effects whose degree of variation remained constant under all settings, namely the first random slopes, $\tau_{12}^2$, the second random intercepts, $\tau^2_{21}$, and the second random slopes $\tau^2_{22}$. Overall, we see clear evidence for a random effect when the true sample standard deviation is also positive ($\tau_{12}^2$ and $\tau_{21}^2$) and evidence for a fixed effect when the sample standard deviation is also approximately zero ($\tau_{22}^2$) (which is only available using the Bayesian estimates).

\begin{table}[ht]
\centering
\begin{tabular}{llcccccccc}
\hline
&& \multicolumn{8}{c}{$\log(\text{EBF}_{01})$}\\
  \hline
 & &\multicolumn{4}{c}{Bayesian estimates} & &\multicolumn{3}{c}{Classical estimates}\\
 \cmidrule{3-6} \cmidrule{8-10}
$J$ & $n$ & $\tau^2_{11}$ & $\tau^2_{12}$ & $\tau^2_{21}$ & $\tau^2_{22}$
   & &$\tau^2_{11}$  & $\tau^2_{12}$ & $\tau^2_{21}$ \\
  \hline
10 & 10 & 4.0 & -1.4e2 & -1.8e2 & 4.2 & & -0.9 & -2.2e2 & -2.1e2 \\
10 & 30 & 3.6 & -6.6e2 & -7.0e2 & 4.8 & & -0.2 & -6.8e2 & -6.4e2 \\
10 & 100 & 4.3 & -2.0e3 & -2.2e3 & 6.7 & & -3.2 & -2.2e3 & -2.1e3 \\
30 & 10 & 6.3 & -6.2e2 & -6.8e2 & 5.9 & & -2.6 & -6.5e2 & -6.5e2 \\
30 & 30 & 7.1 & -2.2e3 & -2.1e3 & 5.6 & & - & -2.1e3 & -2.1e3 \\
30 & 100 & 6.8 & -7.6e3  & -7.5e3 & 4.8 & & -14 & -7.0e3 & -6.5e3 \\
100 & 10 & 16.9 & -2.4e3 & -2.3e3 & 4.9 & & -6.9 & -2.2e3 & -2.3e3 \\
100 & 30 & 16.9 & -8.6e3 & -8.0e3 & 5.0 & & -2.3 & -7.1e3 & -7.0e3 \\
100 & 100 & 21.5 & -3.0e4 & -2.5e4 & 5.0 & & -3.8 & -2.4e4 & -2.2e4 \\
   \hline
\end{tabular}
\caption{Logarithm of the empirical Bayes factors (EBF) between $M_0:\tau_{11}^2=0$ versus $M_1:\tau_{11}^2>0$ in the synthetic data study. To get the EBFs, the sample standard deviations were equal to $\hat{\tau}_{11}=0$ (Bayesian fit), $\hat{\tau}_{11}=0.03$ (classical fit), $\hat{\tau}_{12}=0.5$, $\hat{\tau}_{21}=0.5$, and $\hat{\tau}_{22}=0.01$ (for both approaches). For $J=30$, $n=30$, $\hat{\tau}_{11}$, the classical fit was singular and thus omitted (-).}
\label{tab:simulation}
\end{table}

\section{Empirical applications}\label{sec:application}
In this section, we present five applications to illustrate the use and applicability of our method for various testing problems. In particular, we consider testing the presence of crossed random effects in generalized linear mixed effects models \citep{verbeke2012linear}, spatial random effects \citep{spiegelhalter2002bayesian}, crossed random effects in dynamic structural equation models \citep{asparouhov2018dynamic, mcneish2021measurement}, random intercepts in cross-lagged panel models \citep{hamaker2015critique}, and nonlinear effects in Gaussian process models \citep{williams2006gaussian}. In all these illustrations, we first provide the full model Model M$_1$ and test it against different constrained models with certain random effects are assumed fixed. Moreover, the results are compared with the results of existing approaches which have been reported in the literature. All applications were run on a personal laptop with processor Intel(R) Core(TM) i7 2.80GHz and 16GB RAM.

\subsection{Testing crossed random effects in glmms}

\par
Generalized linear mixed-effect models (glmms) are ubiquitous in applied statistical practice \citep{mccullagh2019generalized, gelman2006data}. They allow researchers to perform analysis of data nested in groups, accounting for the variability among those groups. In particular, crossed random effects are used when different potential sources of heterogeneity are present in different dimensions in a nonnested manner \citep{verbeke2012linear}. Thus, these models provide researchers with simple and effective tools to account for the potential variability of different grouping factors. In this illustration, we conduct an analysis to show how the empirical Bayes factor can be used to test crossed random effects.      

\par
\cite{roback2021beyond} discuss an application, previously presented by \cite{anderson2009officiating}, where they sought evidence that officials from NCAA men's college basketball tend to even out fouls during the game. The response variable, ${y}_{ijhg}$ indicates whether a foul was called favoring the home team, which is modeled using a logistic regression model (1 if it favored the home team and 0 otherwise), where $i$ is the call, $j$ is the game, $h$ is the home team, and $g$ is the visitor team indicators. In total, there were 340 games and 4972 foul calls in the data set. It was hypothesized that this variable can be explained as a function of the foul differential which was defined by the difference between the number of foul calls received by the home and visitor teams. The following logistic random effects model was specified

\bigskip


\noindent
\begin{equation}
\begin{gathered}
\label{eq:roback}
    y_{ijhg} \sim \text{Bernoulli}\Big(\frac{1}{1 + e^{-\mu_{ijhg}(\bm{\theta}_{1j}, \bm{\theta}_{2h}, \bm{\theta}_{3g})}}\Big)\\
    \mu_{ijhg}(\bm\theta_{1j},\bm\theta_{2h},\bm\theta_{3g})) =
    \alpha_0 + \theta_{1j} + \theta_{2h} + \theta_{3g} + (\beta_0 + \theta_{1j^*} + \theta_{2h^*}
    + \theta_{3g^*})x_{ijhg}\\
    \bm{\theta}_{1j} \sim \mathcal{N}(\bm{0}, \bm{\Psi}_1)\\
    \bm{\theta}_{2h} \sim \mathcal{N}(\bm{0}, \bm{\Psi}_2)\\
    \bm{\theta}_{3g} \sim \mathcal{N}(\bm{0}, \bm{\Psi}_3)
\end{gathered}
\end{equation}

\noindent
where $\bm{\Psi}_1 = \text{diag}(\tau^2_{1},\tau^2_{1^*})$, $\bm{\Psi}_2 = \text{diag}(\tau^2_{2},\tau^2_{2^*})$ and $\bm{\Psi}_3 = \text{diag}(\tau^2_{3},\tau^2_{3^*})$. Also, $x_{ijhg}$ is the foul differential for call $i = 1, \dots, 4972$, game $j = 1, \dots, 340$, home team $h = 1, \dots, 39$ and visitor $g = 1, \dots, 39$. The $*$ on $i^*$, $h^*$ and $g^*$ is just used to differentiate between random intercept and random effect for foul differential. Finally, following our standard notation, $\bm{\phi} = (\alpha_0, \beta_0)$.







\par
We computed EBFs to assess the need for including of each of these three random intercepts and three random slopes. \cite{roback2021beyond} acknowledged the problem with the asymptotic distribution of the likelihood ratio test statistic, and thus they carried out a parametric bootstrap. Besides having to fit additional models, the parametric bootstrap is a computationally intensive method and requires a large number of simulations to approximate the bootstrap distribution. Therefore, \cite{roback2021beyond} only fit the model described in Equation \eqref{eq:roback} and test it against a model that did not contain foul differential as random slope for all the three factors. After running 1000 simulations, the final p value was equal to 0.454, from which they concluded that foul differential should be used as fixed slopes. Running this simulation on the machine we used took approximately 6 hours.

\par
In contrast, our empirical Bayes factor approach only requires to fit the model described in Equation \eqref{eq:roback}. We fitted the model using \textit{\textbf{lme4}} \citep{bates2015package} because \cite{roback2021beyond} also used a classical fit of the model. From the fitted model, we can test all the random effects separately and evaluate which ones should be included in a final model. By doing so, via Equation \eqref{eq:SD_bf}, we avoid fitting additional models and the results are virtually instantaneous, as opposed to the 6 hours required to run the parametric bootstrap. Table \ref{tab:basketball} shows the results of the logarithm of the Empirical Bayes Factor for all random effects. From this table we can conclude that there is no considerable evidence that foul differential should be included as a random effect in the model. Moreover, there is only slight evidence for the inclusion of the random intercept at the visitor level.
The parametric bootstrap also arrived at the conclusion that only random intercepts should be included in the model. Thus, foul differential should be used as a fixed effect in the model.

\begin{table}[ht]
\centering
\begin{tabular}{lcc}
  \hline
  \multicolumn{3}{c}{Log of EBF - Basketball data} \\ 
  \hline
           & foul differential & intercept \\ 
  game     &  -0.0022   & -16.8118 \\ 
  home team & -0.2386    & -7.7807 \\ 
  visitor &  0.0000     & -1.2157 \\ 
   \hline
\end{tabular}
\caption{Logarithm of the empirical Bayes factors (EBF) for testing each of the 6 random effects to be fixed against the full random effects model for the basketball data application from \cite{roback2021beyond}.}
\label{tab:basketball}
\end{table}

\subsection{Testing spatial random effects}

\par
Spatial random effects models are frequently employed in spatial statistics and analysis \citep{kang2011bayesian, baltagi2013generalized}. They serve to address spatial autocorrelation, which occurs when spatially close observations exhibit greater similarity than those further apart. These effects prove invaluable in the analysis of spatially correlated datasets, including geographical data and data obtained from spatially distributed entities such as households, census tracts, or grid cells.

\par
\cite{spiegelhalter2002bayesian} explored the data set on cancer lip rates in 56 counties in Scotland. The data contains information on the observed ($y_{j}$) and expected ($x_{j}$) number of lip cancer cases in each county. They assume the observed counts to follow a Poisson distribution, where $e^{\mu_{j}}$ denotes the true area-specific risk of lip cancer, for area $j = 1, \dots, 56$. The model is defined as

\bigskip


\noindent
\begin{equation}
    \begin{gathered}
    \label{eq:spatial_random}
        y_{j} \sim \text{Poisson}\Big( \exp(\mu_{j}(\bm\theta)) x_{j}\Big)\\
        \mu_j(\bm\theta) = \alpha_{0} + \theta_{j1} + \theta_{j2}\\
        \theta_{j1} \sim \mathcal{N}(0, \tau^2_{\theta_1})\\
        \theta_{j2} | \bm{\theta}_{-j2} \sim \mathcal{N}\Bigg(\frac{1}{L_n} \sum_{\ell=1}^{L_n} \theta_{j2}, \frac{\tau^{2}_{\bm{\theta}_2}}{L_n} \Bigg),
    \end{gathered}
\end{equation}

\noindent
where $\alpha_{0}$ is a fixed intercept, $\theta_{n1}$ are exchangeable random effects, $\theta_{n2}$ are spacial random effects and $\bm{\theta}_{-j2}$ is a vector containing all elements of $\bm{\theta}_2$ except the $j$-th component. Moreover, $L_{n}$ is the number of counties adjacent to county $j$, $\tau^{2}_{\bm{\theta}_2}$ is a random-effect variance parameter, and $\bm{\phi} = (\alpha_{0})$. The distribution of $\bm{\theta}_2 = (\theta_{1,2}, \dots, \theta_{56,2})$ implies a structured covariance matrix
.
To see this, we can write the joint distribution of $\bm{\theta}_2$ as

\begin{equation}
    \bm{\theta}_2 = \bm{W} \bm{\theta}_2 + \mathcal{N}(\bm{0}, \tau^{2}_{\bm{\theta}_2} \text{diag}(1/L_{1}, \dots, 1/L_{56})),
\end{equation}

\noindent
where $\bm{W}$ is a weight matrix and $\text{diag}(1/J_{1}, \dots, 1/J_{56})$ is a diagonal matrix with entries corresponding to the number of adjacent counties to each county $j$. After some probability calculus, the joint distribution of $\bm{\theta}_2$ can be written as

\begin{equation}
\label{eq:struct_model}
    \bm{\theta}_2 \sim \mathcal{N}\Big(\bm{0}, \tau^{2}_{\bm{\theta}_2} {(\bm{I} - \bm{W})}^{-1} \bm{B} {({(\bm{I} - \bm{W})}^{-1})}^{T} \Big),
\end{equation}

\noindent
where $\bm{I}$ is an identity matrix with dimensions $56 \times 56$ and $\bm{B}$ is $\text{diag}(1/L_{1}, \dots, 1/L_{56})$. Thus, the model has two random effects where the first has a simple diagonal covariance structure $\tau^2_{\theta_1}\textbf{I}$ and the second has a structured covariance structure as in Equation \eqref{eq:struct_model}.

\par
\cite{spiegelhalter2002bayesian} fitted all four random effects models (by including/excluding the first and second random effect) and computed the respective DICs. The model where both random effects were omitted resulted in the poorest fit while all other three random effects models resulted in an acceptable fit. In the end, the random effects model where the second (spatial) random effect was included and the first random effect was excluded was preferred based the number of effective parameters and the fit of these three models.


\par
To compute the empirical Bayes factors, only the full random effects model in Equation \eqref{eq:spatial_random} needed to be fit. The logarithm of the EBF when fixing the first random effect against the full model was equal to $4.7356$ implying the absence of any evidence to use a random effect. Whereas, the logirithm of the EBF when fixing the second (spatial) random effect against the full model was equal to $-2.3595$ implying clear evidence to use a spatial random effect. Based on these results we end up with the same recommendation as \cite{spiegelhalter2002bayesian} where a random effects model is preferred which only includes a spatial random effect.

\subsection{Testing individual and temporal measurement invariance in dynamic structural equation models}

\par
Dynamic structural equation models (DSEMs) are statistical models used in the social sciences to analyze time-series data to understand the development of variables over time and their relationships \citep{asparouhov2018dynamic}. Random effects are particularly useful for this class of models to account heterogeneity caused by temporal dependencies and individual dependencies in addition to latent variables. These models extend the traditional structural equation modeling (SEM) framework by explicitly incorporating temporal dynamics into the model.

\cite{mcneish2021measurement} conducted an analysis using a dynamic structural equation model (DSEM) on data of $n = 1, \dots, 50$ overweight/obese adults with binge eating disorder. These subjects were observed during a 28-day period at 4-hour intervals, which amounts to $t = 1, \dots, 152$ intervals. Three items associated with perseverance were used in their analysis. The full DSEM is formulated as follows

\noindent
Model M$_1$:

\begin{equation}
\begin{gathered}
\label{eq:dsem_time_person_rnd_eff}
    \bm{y}_{nt} \sim \mathcal{N}(\bm{\nu}_{nt} + \bm{\Lambda}_{nt} \bm{\eta}_{nt}, \bm{\Sigma})\\
    {\eta}_{nt} = {\eta_w}+{\theta}_{n 1} + {\theta}_{t 4}\\
    \bm{\nu}_{nt} = \bm{\alpha}_{\nu} + \bm{\theta}_{n 2} + \bm{\theta}_{t 5}\\
    \bm{\Lambda}_{nt} = \bm{\alpha}_{\Lambda} + \bm{\theta}_{n 3} + \bm{\theta}_{t 6}\\
    \bm{\theta}_{n} \sim \mathcal{N}(\bm{0}, \bm{\Psi}_{\theta_n})\\
    \bm{\theta}_{t} \sim \mathcal{N}(\bm{0}, \bm{\Psi_{\theta_t}}),\\
\end{gathered}
\end{equation}


\noindent
where {$\eta_w\sim N(0,1)$}, $\bm{y}_{nt}$ is a $3 \times 1$ vector containing the responses of subject $n$ at time $t$. Moreover, $\bm{\nu}_{nt}$ is a $3 \times 1$ vector of intercepts, $\bm{\Lambda}_{nt}$ is a $3 \times 3$ diagonal matrix with factor loadings, $\bm{\eta}_{nt}$ is a $3 \times 1$ vector of latent variables, $\bm{\Sigma}$ is a diagonal matrix, and $\bm{I}$ is an identity matrix. Moreover, $\bm{\alpha}_{\nu}$ and $\bm{\alpha}_{\Lambda}$ are vectors containing intercepts for the general model intercept and the factor loading matrix. Moreover, $\theta_{n 1}$, $\bm{\theta}_{n 2}$, $\bm{\theta}_{n 3}$ are random effects clustering subjects, with $\bm{\theta}_{n 2}$ and $\bm{\theta}_{n 3}$ being $3 \times 1$ vectors. Thus $\bm{\theta}_{n} = (\theta_{n1}, \theta_{n21}, \theta_{n22}, \theta_{n23}, \theta_{n31}, \theta_{n32}, \theta_{n33})'$ and $\bm{\Psi}_{\theta_n} = \text{diag}(\tau^2_{n1}, \tau^2_{n21}, \tau^2_{n22}, \tau^2_{n23}, \tau^2_{n31}, \tau^2_{n32}, \tau^2_{n33})$. Similarly, $\theta_{t 4}$, $\bm{\theta}_{t 5}$, $\bm{\theta}_{t 6}$ are random effects clustering time, with $\bm{\theta}_{t 5}$ and $\bm{\theta}_{t 6}$ being $3 \times 1$ vectors. Hence, $\bm{\theta}_{t} = (\theta_{t1}, \theta_{t21}, \theta_{t22}, \theta_{t23}, \theta_{t31}, \theta_{t32}, \theta_{t33})'$ and $\bm{\Psi}_{\theta_t} = \text{diag}(\tau^2_{t1}, \tau^2_{t21}, \tau^2_{t22}, \tau^2_{t23}, \tau^2_{t31}, \tau^2_{t32}, \tau^2_{t33})$. Therefore, by testing the random effects of the intercepts and loadings across individuals and across measurement occasions (time), we can check individual and temporal measurement invariance.

\par



\par
\cite{mcneish2021measurement} eyeballed the random-effect variance estimates and concluded that the time-level random-effect variance of the factor loadings was equal to zero, whereas the individual-level random-effect variance was about 0.7 (see Table 2 in \cite{mcneish2021measurement}). Therefore, factor loading random effect for time level should not be included in the model. Table \ref{tab:bingeeating} shows the logarithm of the EBFs for this illustration. Given that the EBFs for the time level of the factor loadings (last 3 rows) are all positive, we would reach the same conclusion, without running into the known pitfalls of eyeballing (see Section \ref{sec:alternatives}). Therefore, the $\bm{\theta}_{t}$ in the fourth row of Equation \ref{eq:dsem_time_person_rnd_eff} should disappear and the simpler model should be preferred.

\begin{table}[ht]
\centering
\begin{tabular}{lcc}
  \hline
   & person level & time level \\ 
   \hline
  intercept item 1 & -384.15 & -1117.82 \\ 
  intercept item 2 & -502.39 & -1433.04 \\ 
  intercept item 3 & -277.78 & -590.23 \\ 
  loading item 1 & -1294.08 & 28.80 \\ 
  loading item 2 & -2121.06 & 30.50 \\ 
  loading item 3 & -1569.98 & 31.10 \\  
   \hline
\end{tabular}
\caption{Logarithm of the Empirical Bayes Factor for the binge eating disorder data application from \cite{mcneish2021measurement}.}
\label{tab:bingeeating}
\end{table}


\subsection{Testing random intercepts in cross-lagged panel models}

\par
Cross-lagged panel models are widely used for analyzing reciprocal influences between different psychological constructs over time in longitudinal panel data. In this illustration, we compare two models: the standard Cross-Lagged Panel Model (CLPM) and its extension, the Random Intercept Cross-Lagged Panel Model (RI-CLPM), introduced by \cite{hamaker2015critique}. 

\par
The RI-CLPM is defined as

\noindent
\begin{equation}
\begin{gathered}
    x_{nt} = \mu_{t} + \theta_{n1} + p_{nt}\\
    y_{nt} = \pi_{t} + \theta_{n2} + q_{nt}\\
    p_{nt}= \alpha_t p_{n, t-1} + \beta_t q_{n, t-1} + u_{nt}\\
    q_{nt}= \delta_t q_{n, t-1} + \gamma_t p_{n, t-1} + \nu_{nt}\\
        \begin{bmatrix}
        \theta_{n1} \\
        \theta_{n2}
    \end{bmatrix} \sim
    \mathcal{N}\Bigg(\bm{0},     \begin{bmatrix}
        \tau^2_{\theta_1} \ \ 0\\
        0 \ \ \tau^2_{\theta_2}
    \end{bmatrix}
    \Bigg)
\end{gathered}
\end{equation}


\noindent
where $x_{nt}$ and $y_{nt}$ represent the two constructs of interest, with $n$ denoting the individual index and $t$ denoting the measurement wave index, $\mu_{t}$ and $\pi_{t}$ are temporal grand means at wave t, $p_{nt}$ and $q_{nt}$ represent individual temporal deviations from these means. 
The autoregressive parameters $\alpha_t$ and $\delta_t$ capture the stability of the constructs over time. The cross-lagged parameters $\beta_t$ and $\gamma_t$ indicate the relations between the constructs (i.e., the effect between $x$ and $y$ over time). Also, $u_{nt}$ and $\nu_{nt}$ are the errors modelled as $(u_{nt},\nu_{nt}) \sim \mathcal{N}(\bm 0, \bm\Sigma)$, with $\bm\Sigma$ an unstructured covariance matrix. In the RI-CLPM, the random intercepts, $\theta_{n1}$ and $\theta_{n2}$, are introduced to differentiate within-person processes from stable between-person differences from a multilevel perspective.





The choice using either CLPM and RI-CLPM influences parameter estimates and, consequently, the conclusions drawn from the data. If stable individual differences exist in the data, applying the CLPM may lead to biased parameter estimates, as noted by \cite{hamaker2015critique}. Conversely, if such differences are negligible, applying the RI-CLPM may overfit the data. According to the Occam's razor principle, also known as the principle of parsimony, the simpler model (i.e., CLPM) should be preferred if the fit is similar to that of a more complex model. This is also related to the bias-variance trade-off. The model will have higher variance and lower predictive power if it is overfitted (due to the unnecessary random intercepts).

\par
To illustrate the application of empirical Bayes factors to test the need to include random intercepts, we use data from \cite{mackinnon2022tutorial}, consisting of responses from 251 participants across five days. The data focuses on two constructs: perfectionistic self-presentation (PSP), measured by three items, and state social anxiety (SSA), measured by seven items. The analysis is based on the (averaged) observed scores of indicators, following the model proposed by \cite{hamaker2011model}. Importantly, note that the empirical Bayes factor can be applied in a straightforward manner to more complex RI-CLPMs with latent variables (as in \cite{mackinnon2022tutorial}) as well as other extensions \citep{mulder2021three}. Here we ignore the uncertainty caused by the measurement model in order to be able to compare the proposed Bayes factor with the chi-bar-square test which seems (to our knowledge) only available for the regular setup without latent variables \citep{kuiper2020}. It is important to note that this significance test, which has a similar purpose as the proposed Bayes factor under this model, can result in inflated type I errors if the null model (i.e., the CLPM) is true, which jeopardizes the reliability of the test outcome. To illustrate this, Appendix \ref{app:chi_squared} shows that the distribution of $p$-values under the CLPM may deviate from a uniform distribution but may actually be more concentrated around 0. This indicates that the test is not able to properly control the type I error rate appropriately, but in fact, rejects the CLPM too often on average. Thereby the test seems to be unable to control the error rate which is the main purpose of this significance test. 

\par
For the empirical data, we ran two analyses: an analysis of the full data and an analysis of a subset of 50 observations. The analysis of the subset was carried out to illustrate a possible difference in conclusion between their $p$-value (which can yield inflated type I error rates) and the proposed EBF. On the full data set, we fitted the RI-CLPM using the package \textit{\textbf{blavaan}} \citep{merkle2015blavaan, merkle2020efficient}. For the full data, the logarithms of the empirical Bayes factors for testing whether the random intercepts $\theta_{n1}$ and $\theta_{n2}$ are constant or not were equal to -1681.14 and -1090.09, implying strong evidence favored the inclusion of random intercepts.

\par
Alternatively, the comparison between the CLPM and RI-CLPM can be assessed using a chi-bar-square test \citep{hamaker2015critique, stoel2006likelihood}. The chi-bar-square distribution is a weighted sum of different chi-square distributions. This test evaluates whether the null model (CLPM) can be rejected in favor of the more complex alternative model (RI-CLPM). A significant $p$-value (i.e., exceeding the threshold that controls the type I error rate) indicates sufficient evidence to reject the null implying to opt for the more complex RI-CLPM. We also ran both models using the \textit{\textbf{lavaan}} package \citep{rosseel2012lavaan} and conducted the chi-bar-square test with `ChiBarSq.DiffTest' \citep{kuiper2020}. We reach the same conclusion that the RI-CLPM is favored with a $p$-value of 0 ($\alpha=.05$).

\par
When conducting the test on the smaller subset, we obtained logarithms of the Bayes factors of $-48.29$ and $1.25$ for the two random intercepts, implying that a random intercept is needed for the first variable (PSP) while there is evidence for fixing the intercept for the second variable (SSA) across observations. When executing the chi-bar-square test, on the other hand, $p$-values of 0 and 0.035 are obtained. This would result in a rejection of both fixed intercepts when using the traditional significance level of .05, and thus a different conclusion. As noted earlier, this difference in behavior can be attributed to (i) this significance test being too liberal resulting in inflated type I errors (Appendix \ref{app:chi_squared}), and (ii) the general tendency of significance tests to overestimate the evidence against a simpler null model \citep{sellke2001calibration,wagenmakers2007practical}.

\subsection{Testing nonlinear effects in Gaussian process regression models}

Nonlinear statistical models are useful to study nonlinear mechanisms between variables which are ubiquitous in scientific practice. Gaussian processes are very flexible to construct nonlinear statistical models for this purpose \citep{williams2006gaussian}. An important question when constructing such models is whether the effects are indeed nonlinear. This is important to address to avoid the need for an overly complex nonlinear model, as (generalized) linear models are easier to interpret and easier to draw inferences about \citep{mulder2023bayesian}. As Gaussian processes can be viewed as a random effects approach with a specific structured covariance structure, we can use the proposed empirical Bayes factor for testing nonlinear effects under nonlinear Gaussian process models. 

\cite{gelman2006data} presented an empirical application to study the relationship between the mother's IQ and the test score of her child. Data were also available on whether the mothers finished their high school. The following nonlinear Gaussian process model is considered where a nonlinear effect is modeled of the mother's IQ, $x_{1}$, and the test score of her child, $y$, and a nonlinear interaction effect of the mother's IQ and whether she finished high school, $x_{2n}$ (where $x_{2}$ equals $x_{1}$ where she finished high school, or else zero):
\begin{eqnarray*}
\bm{y} &=& \bm{f}_1(\bm{x}_1) + \bm{f}_2(\bm{x}_2) + \bm{\epsilon}\\
\bm{f}_1(\bm{x}_1) &\sim& \mathcal{GP}(\textbf{0},\bm k_1(\bm{x}_1,\bm{x}_1'|\tau_1,\lambda_1))\\
\bm{f}_2(\bm{x}_2) &\sim& \mathcal{GP}(\textbf{0},\bm k_2(\bm{x}_2,\bm{x}_2'|\tau_2,\lambda_2))\\
\bm{\epsilon}&\sim & N(\textbf{0},\sigma^2\textbf{I}),
\end{eqnarray*}
where a squared exponential kernel is used for the nonlinear effects
\[
\bm k_p(\bm{x}_p,\bm{x}_p'|\tau_p,\lambda_p) = \tau_p^2\exp\{-\tfrac{||\bm{x}_1-\bm{x}_1'||^2}{2\lambda_p^2}\},
\]
where $\lambda_p$ is the length scale parameter, which quantifies the smoothness (or wiggliness) of a nonlinear effect, and the magnitude parameter $\tau_p$ captures its magnitude, for $p=1$ or 2. For a vector of finite length, as is the case in statistical practice (here the sample size equals $N=434$), a Gaussian process simply comes down to a multivariate Gaussian distribution (Rasmussen \& Williams, 2006). Therefore, the above model can be viewed as a specific random effects model where the kernels, $\bm k_1$ and $\bm k_2$, correspond to the structured covariance matrices of the two random effects, $\bm{f}_1(\bm{x}_1)$ and $\bm{f}_2(\bm{x}_2)$, having structural parameters $\bm\gamma_p=(\tau_p,\lambda_p)$, for $p=1$ or 2.

In the case, there is no effect of $\bm{x}_1$ on $\bm y$, the function $\bm{f}_1(\bm{x}_1)$ would be a constant, implying that all elements of $\bm{f}_1(\bm{x}_1)$ are equal, i.e., $\mathcal{M}_0:f_1(x_{1n}) = \ldots = f_1(x_{1N})$, which can be tested against an unconstrained alternative $\mathcal{M}_1$. A similar model selection problem can be formulated of whether a nonlinear interaction effect of $\bm x_2$ is present or not. After standardization of the data, the full model was fit using Stan (Stan Development Team, 2024) using virtually flat priors for all free parameters following $N(0,100)$ distributions truncated in $(0,\infty)$. The empirical Bayes factors for $\mathcal{M}_0$ versus $\mathcal{M_1}$ for the main effect of mother's IQ, $x_1$, and the interaction effect of mother's IQ and whether she finished high school, $x_2$, equaled $-75.3$ and $-2.9$, respectively. Based on these results it can be concluded that there is clear support for a nonlinear main effect, and there is some evidence for the existence of a nonlinear interaction effect but the evidence is quite mild. Based on these results, it could be argued to only include the nonlinear main effect to avoid an overly complex nonlinear model.

\section{Discussion and conclusion}\label{sec:discussion}

\par
In empirical research, the collected data are often clustered, e.g., due to hierarchical (multilevel) structures \citep[e.g.,][]{roback2021beyond,spiegelhalter2002bayesian,mcneish2021measurement}. When building statistical models for such data, an important question is whether the implied heterogeneity that is caused by the clustered structure is sufficiently compelling for random effects to be included in the model. Because of the ubiquity of this problem in statistical practice, many approaches have been proposed in the statistical literature to address this question including eyeballing statistical descriptive results, significance tests, cross-validation, information criteria or (regular) Bayes factors \citep{crainiceanu2008likelihood,zhang2008variance,hamaker2011model,mulder2013bayesian,berrar2019cross}
. As discussed in this paper, different methods have different advantages and disadvantages. 

\par
The current paper proposed a novel empirical Bayes factor (EBF) for this testing problem which avoids certain limitations of existing methods. Its two main advantages are its simplicity to compute it (only requiring estimates from a classical or Bayesian model fit) and its flexibility to be able any type of structural heterogeneity under any type of mixed effects model. Moreover, the methodology avoids the need for manual prior specification \cite[by abiding the principle of empirical Bayes][]{casella1985introduction} by using estimated variance components on the second level to serve as prior variances for the random effects on the first level. Computationally expensive methods, such as marginal likelihoods, bootstrap simulations, or cross-validation, are avoided by adopting a Savage-Dickey density ratio together with Gaussian approximations of the posterior. Moreover, only the full mixed effects model needs to be fitted in order to compute the EBFs for all separate random effects which is not the case for existing alternatives. EBFs can be computed either using classical estimates of Bayesian estimates, depending on the preferred methodology of the user. As classical MLEs result in EBFs that are slightly more liberal towards the full random effects model (as was shown in the synthetic data study), a slightly more conservative interpretation may be required.

\par
A synthetic data study showed the general expected behavior of the EBFs in controlled setting while varying the degree of heterogeneity of one random effect, the number of clusters, and the cluster sizes. Moreover, it was shown how the EBF can be used to evaluate the need for cross random effects in generalized linear mixed models (glmms) \citep{roback2021beyond}, spatial random-effects \citep{spiegelhalter2002bayesian}, individual and temporal measurement invariance in dynamic structural equation models \citep{mcneish2021measurement}, random intercepts in cross-lagged panel models \citep{hamaker2011model}, and non-linear effects in Gaussian process regression models \citep{williams2006gaussian}. These applications illustrate the wide range of statistical problems that the proposed methodology can handle. Of course the methodology can also be used in many more modeling scenarios such as multilevel models with more than two levels or nonlinear mixed effects models, to name a few.

\par
As noted earlier, the goal of the proposed testing methodology was not necessarily to present a criterion that is inherently superior to all existing methods as each statistical test has its pros and cons. In the empirical illustrations, we have shown that the proposed empirical Bayes factor can produce similar results as existing methods in specific models and for specific data sets. However, there are also cases where it can reach (slightly) different conclusions than existing methods. The main selling point of the proposed method is that it is extremely easy to apply for testing random effects covariance structures in simple but also in very complex mixed effects models. Thus, the applicability of the proposed EBF goes beyond the applications that were presented here. We leave this for future work.

Finally, it is important to note that the EBF relies on the Savage-Dickey density ratio which has received some attention in the literature including potential caveats \citep{verdinelli1995computing,wetzels2010encompassing,marin2010resolving,heck2019caveat,mulder2022generalization}. In particular, the Savage-Dickey density ratio implies that the prior for the nuisance parameters under the restricted model is equal to the prior under the full model by conditioning on the null restriction that the random effects are zero. Because the random effects are often modelled independently, this would imply that the prior for the nuisance parameter under the full model and restricted model would be equal, which is not problematic. However, alternative choices of the nuisance parameters under the full model could be considered depending on the dependencies across random effects. We leave this topic for future work.


\appendix

\section{Simulation Study: Performance of the Chi-bar-squared test}\label{app:chi_squared}

\par
To explore the performance of $\bar{\chi}^2$ test, we ran a simulation study. In particular, we examined its performance with different sample sizes and waves. Both data generation and data analysis were programmed in R (R Core Team, 2021). The packages \textit{\textbf{mvtnorm}}(Genz \& Bretz, 2009; Genz et al., 2021) and ChiBarSq.DiffTest (Kuiper, 2020) were also involved in addition to \textit{\textbf{lavaan}} and \textit{\textbf{blavaan}}. The package \textit{\textbf{mvtnorm}} was used to get the multivariate normal probabilities when computing the Bayes factor. The $\bar{\chi}^2$ test was performed using ChiBarSq.DiffTest to compare CLPM and RI-CLPM.

\paragraph{Data generation mechanism:} To generate bivariate longitudinal data with different waves, we used the simulateData function in lavaan. We specified RI-CLPM for data generation with $\tau^2 = 0$. Note that there are no random intercepts when $\tau^2 = 0$ for both variables, so we would generate data under the null model, i.e., the CPLM. We varied: (1) the sample size: 50, 100, 250, 500, and 1000; (2) the number of waves: 3, 5, and 10, resulting in 5*3=15 combinations. For each combination, we generated 200 data sets based on the defined true $\tau^2$. In total, we can have 3,000 data sets for the true value of $\tau^2$ to analyze. For simplicity, we kept the following parameters constant in our data generation: (1) cross-lagged effects $c=d=0.2$ (equality constraints across waves); (2) auto-regressive effects $a=b=0.5$ (equality constraints across waves). The specifications of the other parameters can be checked online. Because some models did not converge when we generated data from \textit{\textbf{lavaan}}, we ended up with 2,945 data sets for $\tau^2=0$.

\paragraph{Performance:} We specified RI-CLPM with different waves for data analysis. To evaluate the performance of the $\bar{\chi}^2$ test, we consider the following indicator: p-value from the $\bar{\chi}^2$ test, which compares the chi-square difference between RI-CLPM and CLPM ($\text{H}_0$: CLPM). A significant p-value indicates that there is sufficient evidence to reject the null model CLPM and instead select the more complex model, RI-CLPM, which also implies the presence of random intercepts (for both variables). Since p-values should follow a uniform distribution when $\text{H}_0$ is true, we can inspect the distribution of p-values for data generated through $\tau^2 = 0$. If the distribution is uniform, then the type I error rate is controlled by $\alpha$.

\paragraph{Result - P-values distribution under H0:}
For the chi-bar-square test, The figure below shows distributions of p-values when $\tau_x^2 = \tau_y^2 = 0$. Among the different combinations of sample sizes and waves, none of these follow an approximate uniform distribution. When the p-values have a uniform distribution, the mean should be .5. However, in the chi-bar-square test, the means (red dashed lines) are smaller than .5. This implies an exaggerated Type I error rate, where the probability of falsely rejecting the null hypothesis is larger than the $\alpha$ level set, highlighting the test's failure to control for Type I error. Overall, the p-value is conservative in that it rejects the $\text{H}_0$ when it is true and prefers the more complex model RI-CLPM. The actual type I error exceeds the $\alpha$ level we set. When random intercepts are not required, the RI-CLPM is overfitted, resulting in higher variance and lower statistical power.

\begin{figure}[ht]
    \centering
    \subfloat{\includegraphics[width = 5.5in, height = 2.5in]{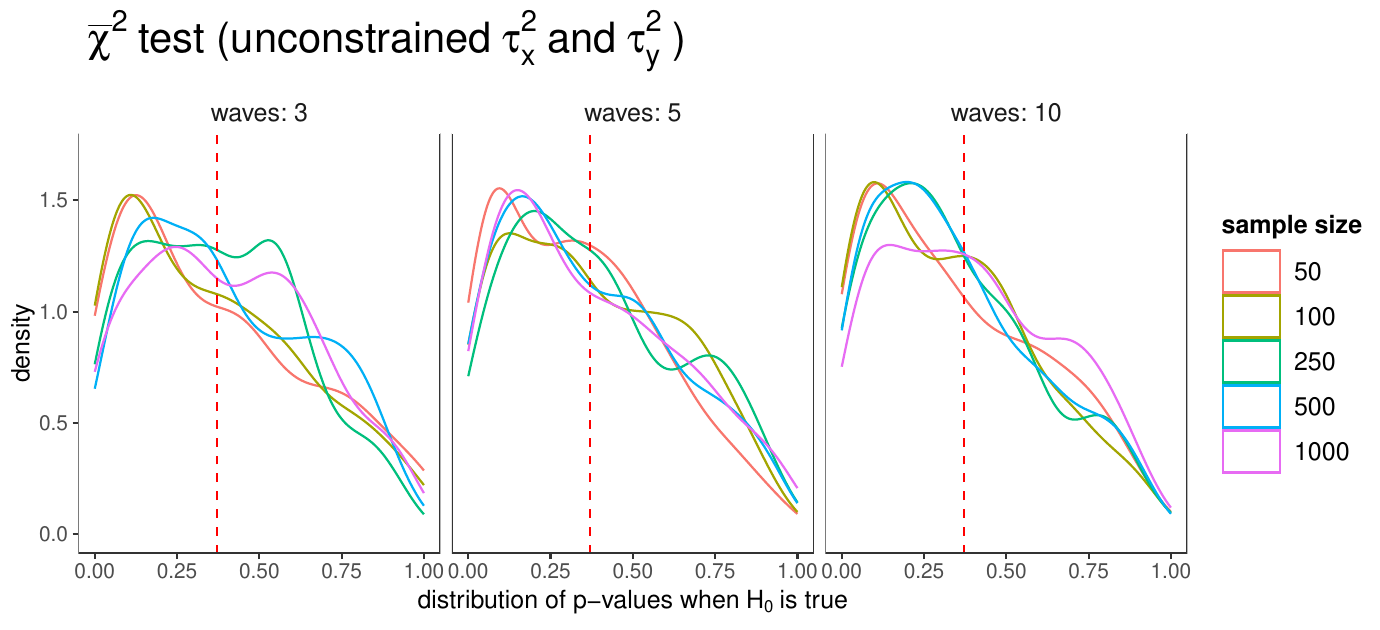}} \\
    \caption*{The p-values distribution from the chi-bar-square test when $\tau_x^2=\tau_y^2=0$. They are classified according to the number of waves and sample size. The dashed lines represent the mean of p-values. The data were analyzed in lavaan without constraints on  $\tau^2$ (where the estimates can be negative). None of the plots show a uniform distribution.}
\end{figure}


\clearpage 
\bibliographystyle{apacite}
\bibliography{references}

\begin{thebibliography}{}

\bibitem [\protect \citeauthoryear {%
Abbott%
\ \BBA {} Machta%
}{%
Abbott%
\ \BBA {} Machta%
}{%
{\protect \APACyear {2022}}%
}]{%
abbott2022far}
\APACinsertmetastar {%
abbott2022far}%
\begin{APACrefauthors}%
Abbott, M\BPBI C.%
\BCBT {}\ \BBA {} Machta, B\BPBI B.%
\end{APACrefauthors}%
\unskip\
\newblock
\APACrefYearMonthDay{2022}{}{}.
\newblock
{\BBOQ}\APACrefatitle {Far from {A}symptopia} {Far from {A}symptopia}.{\BBCQ}
\newblock
\APACjournalVolNumPages{arXiv preprint arXiv:2205.03343}{}{}{}.
\PrintBackRefs{\CurrentBib}

\bibitem [\protect \citeauthoryear {%
Akaike%
}{%
Akaike%
}{%
{\protect \APACyear {1998}}%
}]{%
akaike1998information}
\APACinsertmetastar {%
akaike1998information}%
\begin{APACrefauthors}%
Akaike, H.%
\end{APACrefauthors}%
\unskip\
\newblock
\APACrefYearMonthDay{1998}{}{}.
\newblock
{\BBOQ}\APACrefatitle {Information theory and an extension of the maximum
  likelihood principle} {Information theory and an extension of the maximum
  likelihood principle}.{\BBCQ}
\newblock
\APACjournalVolNumPages{Selected papers of hirotugu akaike}{}{}{199--213}.
\PrintBackRefs{\CurrentBib}

\bibitem [\protect \citeauthoryear {%
Alsakka%
\ \BBA {} Ap~Gwilym%
}{%
Alsakka%
\ \BBA {} Ap~Gwilym%
}{%
{\protect \APACyear {2010}}%
}]{%
alsakka2010random}
\APACinsertmetastar {%
alsakka2010random}%
\begin{APACrefauthors}%
Alsakka, R.%
\BCBT {}\ \BBA {} Ap~Gwilym, O.%
\end{APACrefauthors}%
\unskip\
\newblock
\APACrefYearMonthDay{2010}{}{}.
\newblock
{\BBOQ}\APACrefatitle {A random effects ordered probit model for rating
  migrations} {A random effects ordered probit model for rating
  migrations}.{\BBCQ}
\newblock
\APACjournalVolNumPages{Finance Research Letters}{7}{3}{140--147}.
\PrintBackRefs{\CurrentBib}

\bibitem [\protect \citeauthoryear {%
Anderson%
\ \BBA {} Pierce%
}{%
Anderson%
\ \BBA {} Pierce%
}{%
{\protect \APACyear {2009}}%
}]{%
anderson2009officiating}
\APACinsertmetastar {%
anderson2009officiating}%
\begin{APACrefauthors}%
Anderson%
\BCBT {}\ \BBA {} Pierce, D.%
\end{APACrefauthors}%
\unskip\
\newblock
\APACrefYearMonthDay{2009}{}{}.
\newblock
{\BBOQ}\APACrefatitle {Officiating bias: The effect of foul differential on
  foul calls in NCAA basketball} {Officiating bias: The effect of foul
  differential on foul calls in ncaa basketball}.{\BBCQ}
\newblock
\APACjournalVolNumPages{Journal of sports sciences}{27}{7}{687--694}.
\PrintBackRefs{\CurrentBib}

\bibitem [\protect \citeauthoryear {%
D.~Anderson%
\ \BBA {} Burnham%
}{%
D.~Anderson%
\ \BBA {} Burnham%
}{%
{\protect \APACyear {2002}}%
}]{%
anderson2002avoiding}
\APACinsertmetastar {%
anderson2002avoiding}%
\begin{APACrefauthors}%
Anderson, D.%
\BCBT {}\ \BBA {} Burnham, K\BPBI P.%
\end{APACrefauthors}%
\unskip\
\newblock
\APACrefYearMonthDay{2002}{}{}.
\newblock
{\BBOQ}\APACrefatitle {Avoiding pitfalls when using information-theoretic
  methods} {Avoiding pitfalls when using information-theoretic methods}.{\BBCQ}
\newblock
\APACjournalVolNumPages{The Journal of wildlife management}{}{}{912--918}.
\PrintBackRefs{\CurrentBib}

\bibitem [\protect \citeauthoryear {%
Anguita%
, Ghelardoni%
, Ghio%
, Oneto%
\BCBL {}\ \BBA {} Ridella%
}{%
Anguita%
\ \protect \BOthers {.}}{%
{\protect \APACyear {2012}}%
}]{%
anguita2012k}
\APACinsertmetastar {%
anguita2012k}%
\begin{APACrefauthors}%
Anguita, D.%
, Ghelardoni, L.%
, Ghio, A.%
, Oneto, L.%
\BCBL {}\ \BBA {} Ridella, S.%
\end{APACrefauthors}%
\unskip\
\newblock
\APACrefYearMonthDay{2012}{}{}.
\newblock
{\BBOQ}\APACrefatitle {The'K'in K-fold Cross Validation.} {The'k'in k-fold
  cross validation.}{\BBCQ}
\newblock
\BIn{} \APACrefbtitle {ESANN} {Esann}\ (\BPGS\ 441--446).
\PrintBackRefs{\CurrentBib}

\bibitem [\protect \citeauthoryear {%
Asparouhov%
, Hamaker%
\BCBL {}\ \BBA {} Muth{\'e}n%
}{%
Asparouhov%
\ \protect \BOthers {.}}{%
{\protect \APACyear {2018}}%
}]{%
asparouhov2018dynamic}
\APACinsertmetastar {%
asparouhov2018dynamic}%
\begin{APACrefauthors}%
Asparouhov, T.%
, Hamaker, E\BPBI L.%
\BCBL {}\ \BBA {} Muth{\'e}n, B.%
\end{APACrefauthors}%
\unskip\
\newblock
\APACrefYearMonthDay{2018}{}{}.
\newblock
{\BBOQ}\APACrefatitle {Dynamic structural equation models} {Dynamic structural
  equation models}.{\BBCQ}
\newblock
\APACjournalVolNumPages{Structural equation modeling: a multidisciplinary
  journal}{25}{3}{359--388}.
\PrintBackRefs{\CurrentBib}

\bibitem [\protect \citeauthoryear {%
Baltagi%
, Egger%
\BCBL {}\ \BBA {} Pfaffermayr%
}{%
Baltagi%
\ \protect \BOthers {.}}{%
{\protect \APACyear {2013}}%
}]{%
baltagi2013generalized}
\APACinsertmetastar {%
baltagi2013generalized}%
\begin{APACrefauthors}%
Baltagi, B\BPBI H.%
, Egger, P.%
\BCBL {}\ \BBA {} Pfaffermayr, M.%
\end{APACrefauthors}%
\unskip\
\newblock
\APACrefYearMonthDay{2013}{}{}.
\newblock
{\BBOQ}\APACrefatitle {A generalized spatial panel data model with random
  effects} {A generalized spatial panel data model with random effects}.{\BBCQ}
\newblock
\APACjournalVolNumPages{Econometric reviews}{32}{5-6}{650--685}.
\PrintBackRefs{\CurrentBib}

\bibitem [\protect \citeauthoryear {%
Bates%
\ \protect \BOthers {.}}{%
Bates%
\ \protect \BOthers {.}}{%
{\protect \APACyear {2015}}%
}]{%
bates2015package}
\APACinsertmetastar {%
bates2015package}%
\begin{APACrefauthors}%
Bates, D.%
, Maechler, M.%
, Bolker, B.%
, Walker, S.%
, Christensen, R\BPBI H\BPBI B.%
, Singmann, H.%
\BDBL {}Bolker, M\BPBI B.%
\end{APACrefauthors}%
\unskip\
\newblock
\APACrefYearMonthDay{2015}{}{}.
\newblock
{\BBOQ}\APACrefatitle {Package ‘lme4’} {Package ‘lme4’}.{\BBCQ}
\newblock
\APACjournalVolNumPages{convergence}{12}{1}{2}.
\PrintBackRefs{\CurrentBib}

\bibitem [\protect \citeauthoryear {%
Berrar%
}{%
Berrar%
}{%
{\protect \APACyear {2019}}%
}]{%
berrar2019cross}
\APACinsertmetastar {%
berrar2019cross}%
\begin{APACrefauthors}%
Berrar, D.%
\end{APACrefauthors}%
\unskip\
\newblock
\APACrefYearMonthDay{2019}{}{}.
\newblock
\APACrefbtitle {Cross-Validation.} {Cross-validation.}
\PrintBackRefs{\CurrentBib}

\bibitem [\protect \citeauthoryear {%
Borenstein%
, Hedges%
, Higgins%
\BCBL {}\ \BBA {} Rothstein%
}{%
Borenstein%
\ \protect \BOthers {.}}{%
{\protect \APACyear {2010}}%
}]{%
borenstein2010basic}
\APACinsertmetastar {%
borenstein2010basic}%
\begin{APACrefauthors}%
Borenstein, M.%
, Hedges, L\BPBI V.%
, Higgins, J\BPBI P.%
\BCBL {}\ \BBA {} Rothstein, H\BPBI R.%
\end{APACrefauthors}%
\unskip\
\newblock
\APACrefYearMonthDay{2010}{}{}.
\newblock
{\BBOQ}\APACrefatitle {A basic introduction to fixed-effect and random-effects
  models for meta-analysis} {A basic introduction to fixed-effect and
  random-effects models for meta-analysis}.{\BBCQ}
\newblock
\APACjournalVolNumPages{Research synthesis methods}{1}{2}{97--111}.
\PrintBackRefs{\CurrentBib}

\bibitem [\protect \citeauthoryear {%
B{\"u}rkner%
}{%
B{\"u}rkner%
}{%
{\protect \APACyear {2017}}%
}]{%
burkner2017brms}
\APACinsertmetastar {%
burkner2017brms}%
\begin{APACrefauthors}%
B{\"u}rkner, P\BHBI C.%
\end{APACrefauthors}%
\unskip\
\newblock
\APACrefYearMonthDay{2017}{}{}.
\newblock
{\BBOQ}\APACrefatitle {brms: An R package for Bayesian multilevel models using
  Stan} {brms: An r package for bayesian multilevel models using stan}.{\BBCQ}
\newblock
\APACjournalVolNumPages{Journal of statistical software}{80}{}{1--28}.
\PrintBackRefs{\CurrentBib}

\bibitem [\protect \citeauthoryear {%
Casella%
}{%
Casella%
}{%
{\protect \APACyear {1985}}%
}]{%
casella1985introduction}
\APACinsertmetastar {%
casella1985introduction}%
\begin{APACrefauthors}%
Casella, G.%
\end{APACrefauthors}%
\unskip\
\newblock
\APACrefYearMonthDay{1985}{}{}.
\newblock
{\BBOQ}\APACrefatitle {An introduction to empirical {B}ayes data analysis} {An
  introduction to empirical {B}ayes data analysis}.{\BBCQ}
\newblock
\APACjournalVolNumPages{The American Statistician}{39}{2}{83--87}.
\PrintBackRefs{\CurrentBib}

\bibitem [\protect \citeauthoryear {%
Chung%
, Rabe-Hesketh%
, Dorie%
, Gelman%
\BCBL {}\ \BBA {} Liu%
}{%
Chung%
\ \protect \BOthers {.}}{%
{\protect \APACyear {2013}}%
}]{%
chung2013variance}
\APACinsertmetastar {%
chung2013variance}%
\begin{APACrefauthors}%
Chung, Y.%
, Rabe-Hesketh, S.%
, Dorie, V.%
, Gelman, A.%
\BCBL {}\ \BBA {} Liu, J.%
\end{APACrefauthors}%
\unskip\
\newblock
\APACrefYearMonthDay{2013}{}{}.
\newblock
{\BBOQ}\APACrefatitle {A nondegenerate penalized likelihood estimator for
  variance parameters in multilevel models} {A nondegenerate penalized
  likelihood estimator for variance parameters in multilevel models}.{\BBCQ}
\newblock
\APACjournalVolNumPages{Psychometrika}{78}{4}{685--709}.
\newblock
\begin{APACrefDOI} \doi{10.1007/s11336-013-9328-2} \end{APACrefDOI}
\PrintBackRefs{\CurrentBib}

\bibitem [\protect \citeauthoryear {%
Cleophas%
\ \BBA {} Zwinderman%
}{%
Cleophas%
\ \BBA {} Zwinderman%
}{%
{\protect \APACyear {2008}}%
}]{%
cleophas2008random}
\APACinsertmetastar {%
cleophas2008random}%
\begin{APACrefauthors}%
Cleophas, T.%
\BCBT {}\ \BBA {} Zwinderman, A.%
\end{APACrefauthors}%
\unskip\
\newblock
\APACrefYearMonthDay{2008}{}{}.
\newblock
{\BBOQ}\APACrefatitle {Random effects models in clinical research.} {Random
  effects models in clinical research.}{\BBCQ}
\newblock
\APACjournalVolNumPages{International journal of clinical pharmacology and
  therapeutics}{46}{8}{421--427}.
\PrintBackRefs{\CurrentBib}

\bibitem [\protect \citeauthoryear {%
Crainiceanu%
}{%
Crainiceanu%
}{%
{\protect \APACyear {2008}}%
}]{%
crainiceanu2008likelihood}
\APACinsertmetastar {%
crainiceanu2008likelihood}%
\begin{APACrefauthors}%
Crainiceanu, C\BPBI M.%
\end{APACrefauthors}%
\unskip\
\newblock
\APACrefYearMonthDay{2008}{}{}.
\newblock
{\BBOQ}\APACrefatitle {Likelihood ratio testing for zero variance components in
  linear mixed models} {Likelihood ratio testing for zero variance components
  in linear mixed models}.{\BBCQ}
\newblock
\APACjournalVolNumPages{Random effect and latent variable model
  selection}{}{}{3--17}.
\PrintBackRefs{\CurrentBib}

\bibitem [\protect \citeauthoryear {%
Dickey%
}{%
Dickey%
}{%
{\protect \APACyear {1971}}%
}]{%
dickey1971weighted}
\APACinsertmetastar {%
dickey1971weighted}%
\begin{APACrefauthors}%
Dickey, J\BPBI M.%
\end{APACrefauthors}%
\unskip\
\newblock
\APACrefYearMonthDay{1971}{}{}.
\newblock
{\BBOQ}\APACrefatitle {The weighted likelihood ratio, linear hypotheses on
  normal location parameters} {The weighted likelihood ratio, linear hypotheses
  on normal location parameters}.{\BBCQ}
\newblock
\APACjournalVolNumPages{The Annals of Mathematical Statistics}{}{}{204--223}.
\PrintBackRefs{\CurrentBib}

\bibitem [\protect \citeauthoryear {%
Du%
\ \BBA {} Wang%
}{%
Du%
\ \BBA {} Wang%
}{%
{\protect \APACyear {2020}}%
}]{%
du2020testing}
\APACinsertmetastar {%
du2020testing}%
\begin{APACrefauthors}%
Du, H.%
\BCBT {}\ \BBA {} Wang, L.%
\end{APACrefauthors}%
\unskip\
\newblock
\APACrefYearMonthDay{2020}{}{}.
\newblock
{\BBOQ}\APACrefatitle {Testing variance components in linear mixed modeling
  using permutation} {Testing variance components in linear mixed modeling
  using permutation}.{\BBCQ}
\newblock
\APACjournalVolNumPages{Multivariate behavioral research}{55}{1}{120--136}.
\PrintBackRefs{\CurrentBib}

\bibitem [\protect \citeauthoryear {%
Fox%
, Mulder%
\BCBL {}\ \BBA {} Sinharay%
}{%
Fox%
\ \protect \BOthers {.}}{%
{\protect \APACyear {2017}}%
}]{%
fox2017bayes}
\APACinsertmetastar {%
fox2017bayes}%
\begin{APACrefauthors}%
Fox, J\BHBI P.%
, Mulder, J.%
\BCBL {}\ \BBA {} Sinharay, S.%
\end{APACrefauthors}%
\unskip\
\newblock
\APACrefYearMonthDay{2017}{}{}.
\newblock
{\BBOQ}\APACrefatitle {Bayes factor covariance testing in item response models}
  {Bayes factor covariance testing in item response models}.{\BBCQ}
\newblock
\APACjournalVolNumPages{Psychometrika}{82}{}{979--1006}.
\PrintBackRefs{\CurrentBib}

\bibitem [\protect \citeauthoryear {%
Friel%
\ \BBA {} Pettitt%
}{%
Friel%
\ \BBA {} Pettitt%
}{%
{\protect \APACyear {2008}}%
}]{%
friel2008marginal}
\APACinsertmetastar {%
friel2008marginal}%
\begin{APACrefauthors}%
Friel, N.%
\BCBT {}\ \BBA {} Pettitt, A\BPBI N.%
\end{APACrefauthors}%
\unskip\
\newblock
\APACrefYearMonthDay{2008}{}{}.
\newblock
{\BBOQ}\APACrefatitle {Marginal likelihood estimation via power posteriors}
  {Marginal likelihood estimation via power posteriors}.{\BBCQ}
\newblock
\APACjournalVolNumPages{Journal of the Royal Statistical Society: Series B
  (Statistical Methodology)}{70}{3}{589--607}.
\PrintBackRefs{\CurrentBib}

\bibitem [\protect \citeauthoryear {%
Gelman%
\ \BBA {} Hill%
}{%
Gelman%
\ \BBA {} Hill%
}{%
{\protect \APACyear {2006}}%
}]{%
gelman2006data}
\APACinsertmetastar {%
gelman2006data}%
\begin{APACrefauthors}%
Gelman, A.%
\BCBT {}\ \BBA {} Hill, J.%
\end{APACrefauthors}%
\unskip\
\newblock
\APACrefYear{2006}.
\newblock
\APACrefbtitle {Data analysis using regression and multilevel/hierarchical
  models} {Data analysis using regression and multilevel/hierarchical models}.
\newblock
\APACaddressPublisher{}{Cambridge university press}.
\PrintBackRefs{\CurrentBib}

\bibitem [\protect \citeauthoryear {%
Gelman%
, Hwang%
\BCBL {}\ \BBA {} Vehtari%
}{%
Gelman%
\ \protect \BOthers {.}}{%
{\protect \APACyear {2014}}%
}]{%
gelman2014understanding}
\APACinsertmetastar {%
gelman2014understanding}%
\begin{APACrefauthors}%
Gelman, A.%
, Hwang, J.%
\BCBL {}\ \BBA {} Vehtari, A.%
\end{APACrefauthors}%
\unskip\
\newblock
\APACrefYearMonthDay{2014}{}{}.
\newblock
{\BBOQ}\APACrefatitle {Understanding predictive information criteria for
  {B}ayesian models} {Understanding predictive information criteria for
  {B}ayesian models}.{\BBCQ}
\newblock
\APACjournalVolNumPages{Statistics and computing}{24}{6}{997--1016}.
\PrintBackRefs{\CurrentBib}

\bibitem [\protect \citeauthoryear {%
Gelman%
\ \BBA {} Meng%
}{%
Gelman%
\ \BBA {} Meng%
}{%
{\protect \APACyear {1998}}%
}]{%
gelman1998simulating}
\APACinsertmetastar {%
gelman1998simulating}%
\begin{APACrefauthors}%
Gelman, A.%
\BCBT {}\ \BBA {} Meng, X\BHBI L.%
\end{APACrefauthors}%
\unskip\
\newblock
\APACrefYearMonthDay{1998}{}{}.
\newblock
{\BBOQ}\APACrefatitle {Simulating normalizing constants: From importance
  sampling to bridge sampling to path sampling} {Simulating normalizing
  constants: From importance sampling to bridge sampling to path
  sampling}.{\BBCQ}
\newblock
\APACjournalVolNumPages{Statistical science}{}{}{163--185}.
\PrintBackRefs{\CurrentBib}

\bibitem [\protect \citeauthoryear {%
George%
\ \BBA {} Foster%
}{%
George%
\ \BBA {} Foster%
}{%
{\protect \APACyear {2000}}%
}]{%
george2000calibration}
\APACinsertmetastar {%
george2000calibration}%
\begin{APACrefauthors}%
George, E.%
\BCBT {}\ \BBA {} Foster, D\BPBI P.%
\end{APACrefauthors}%
\unskip\
\newblock
\APACrefYearMonthDay{2000}{}{}.
\newblock
{\BBOQ}\APACrefatitle {Calibration and empirical Bayes variable selection}
  {Calibration and empirical bayes variable selection}.{\BBCQ}
\newblock
\APACjournalVolNumPages{Biometrika}{87}{4}{731--747}.
\PrintBackRefs{\CurrentBib}

\bibitem [\protect \citeauthoryear {%
Geweke%
}{%
Geweke%
}{%
{\protect \APACyear {1999}}%
}]{%
geweke1999using}
\APACinsertmetastar {%
geweke1999using}%
\begin{APACrefauthors}%
Geweke, J.%
\end{APACrefauthors}%
\unskip\
\newblock
\APACrefYearMonthDay{1999}{}{}.
\newblock
{\BBOQ}\APACrefatitle {Using simulation methods for Bayesian econometric
  models: inference, development, and communication} {Using simulation methods
  for bayesian econometric models: inference, development, and
  communication}.{\BBCQ}
\newblock
\APACjournalVolNumPages{Econometric reviews}{18}{1}{1--73}.
\PrintBackRefs{\CurrentBib}

\bibitem [\protect \citeauthoryear {%
Gronau%
\ \BBA {} Wagenmakers%
}{%
Gronau%
\ \BBA {} Wagenmakers%
}{%
{\protect \APACyear {2019}}%
}]{%
gronau2019limitations}
\APACinsertmetastar {%
gronau2019limitations}%
\begin{APACrefauthors}%
Gronau, Q\BPBI F.%
\BCBT {}\ \BBA {} Wagenmakers, E\BHBI J.%
\end{APACrefauthors}%
\unskip\
\newblock
\APACrefYearMonthDay{2019}{}{}.
\newblock
{\BBOQ}\APACrefatitle {Limitations of Bayesian leave-one-out cross-validation
  for model selection} {Limitations of bayesian leave-one-out cross-validation
  for model selection}.{\BBCQ}
\newblock
\APACjournalVolNumPages{Computational brain \& behavior}{2}{}{1--11}.
\PrintBackRefs{\CurrentBib}

\bibitem [\protect \citeauthoryear {%
Guo%
, Gabry%
, Goodrich%
\BCBL {}\ \BBA {} Weber%
}{%
Guo%
\ \protect \BOthers {.}}{%
{\protect \APACyear {2020}}%
}]{%
guo2020package}
\APACinsertmetastar {%
guo2020package}%
\begin{APACrefauthors}%
Guo, J.%
, Gabry, J.%
, Goodrich, B.%
\BCBL {}\ \BBA {} Weber, S.%
\end{APACrefauthors}%
\unskip\
\newblock
\APACrefYearMonthDay{2020}{}{}.
\newblock
{\BBOQ}\APACrefatitle {Package ‘rstan’} {Package ‘rstan’}.{\BBCQ}
\newblock
\APACjournalVolNumPages{URL https://cran. r―project.
  org/web/packages/rstan/(2020)}{}{}{}.
\PrintBackRefs{\CurrentBib}

\bibitem [\protect \citeauthoryear {%
Hamaker%
, Kuiper%
\BCBL {}\ \BBA {} Grasman%
}{%
Hamaker%
\ \protect \BOthers {.}}{%
{\protect \APACyear {2015}}%
}]{%
hamaker2015critique}
\APACinsertmetastar {%
hamaker2015critique}%
\begin{APACrefauthors}%
Hamaker, E\BPBI L.%
, Kuiper, R\BPBI M.%
\BCBL {}\ \BBA {} Grasman, R\BPBI P.%
\end{APACrefauthors}%
\unskip\
\newblock
\APACrefYearMonthDay{2015}{}{}.
\newblock
{\BBOQ}\APACrefatitle {A critique of the cross-lagged panel model.} {A critique
  of the cross-lagged panel model.}{\BBCQ}
\newblock
\APACjournalVolNumPages{Psychological methods}{20}{1}{102}.
\PrintBackRefs{\CurrentBib}

\bibitem [\protect \citeauthoryear {%
Hamaker%
, van Hattum%
, Kuiper%
\BCBL {}\ \BBA {} Hoijtink%
}{%
Hamaker%
\ \protect \BOthers {.}}{%
{\protect \APACyear {2011}}%
}]{%
hamaker2011model}
\APACinsertmetastar {%
hamaker2011model}%
\begin{APACrefauthors}%
Hamaker, E\BPBI L.%
, van Hattum, P.%
, Kuiper, R\BPBI M.%
\BCBL {}\ \BBA {} Hoijtink, H.%
\end{APACrefauthors}%
\unskip\
\newblock
\APACrefYearMonthDay{2011}{}{}.
\newblock
{\BBOQ}\APACrefatitle {Model selection based on information criteria in
  multilevel modeling} {Model selection based on information criteria in
  multilevel modeling}.{\BBCQ}
\newblock
\APACjournalVolNumPages{Handbook of advanced multilevel
  analysis}{}{}{231--255}.
\PrintBackRefs{\CurrentBib}

\bibitem [\protect \citeauthoryear {%
Hartwig%
\ \BBA {} Dearing%
}{%
Hartwig%
\ \BBA {} Dearing%
}{%
{\protect \APACyear {1979}}%
}]{%
hartwig1979exploratory}
\APACinsertmetastar {%
hartwig1979exploratory}%
\begin{APACrefauthors}%
Hartwig, F.%
\BCBT {}\ \BBA {} Dearing, B\BPBI E.%
\end{APACrefauthors}%
\unskip\
\newblock
\APACrefYear{1979}.
\newblock
\APACrefbtitle {Exploratory data analysis} {Exploratory data analysis}\
  (\BNUM~16).
\newblock
\APACaddressPublisher{}{Sage}.
\PrintBackRefs{\CurrentBib}

\bibitem [\protect \citeauthoryear {%
Heck%
}{%
Heck%
}{%
{\protect \APACyear {2019}}%
}]{%
heck2019caveat}
\APACinsertmetastar {%
heck2019caveat}%
\begin{APACrefauthors}%
Heck, D\BPBI W.%
\end{APACrefauthors}%
\unskip\
\newblock
\APACrefYearMonthDay{2019}{}{}.
\newblock
{\BBOQ}\APACrefatitle {A caveat on the Savage--Dickey density ratio: The case
  of computing Bayes factors for regression parameters} {A caveat on the
  savage--dickey density ratio: The case of computing bayes factors for
  regression parameters}.{\BBCQ}
\newblock
\APACjournalVolNumPages{British Journal of Mathematical and Statistical
  Psychology}{72}{2}{316--333}.
\PrintBackRefs{\CurrentBib}

\bibitem [\protect \citeauthoryear {%
Ioannidis%
}{%
Ioannidis%
}{%
{\protect \APACyear {2008}}%
}]{%
ioannidis2008interpretation}
\APACinsertmetastar {%
ioannidis2008interpretation}%
\begin{APACrefauthors}%
Ioannidis, J\BPBI P.%
\end{APACrefauthors}%
\unskip\
\newblock
\APACrefYearMonthDay{2008}{}{}.
\newblock
{\BBOQ}\APACrefatitle {Interpretation of tests of heterogeneity and bias in
  meta-analysis} {Interpretation of tests of heterogeneity and bias in
  meta-analysis}.{\BBCQ}
\newblock
\APACjournalVolNumPages{Journal of evaluation in clinical
  practice}{14}{5}{951--957}.
\PrintBackRefs{\CurrentBib}

\bibitem [\protect \citeauthoryear {%
Jeffreys%
}{%
Jeffreys%
}{%
{\protect \APACyear {1961}}%
}]{%
jeffreys1961theory}
\APACinsertmetastar {%
jeffreys1961theory}%
\begin{APACrefauthors}%
Jeffreys, H.%
\end{APACrefauthors}%
\unskip\
\newblock
\APACrefYearMonthDay{1961}{}{}.
\newblock
{\BBOQ}\APACrefatitle {Theory of probability, ed. 3 oxford university press}
  {Theory of probability, ed. 3 oxford university press}.{\BBCQ}
\newblock
\APACjournalVolNumPages{Figure Click here to access/download}{}{}{}.
\PrintBackRefs{\CurrentBib}

\bibitem [\protect \citeauthoryear {%
Kang%
\ \BBA {} Cressie%
}{%
Kang%
\ \BBA {} Cressie%
}{%
{\protect \APACyear {2011}}%
}]{%
kang2011bayesian}
\APACinsertmetastar {%
kang2011bayesian}%
\begin{APACrefauthors}%
Kang, E\BPBI L.%
\BCBT {}\ \BBA {} Cressie, N.%
\end{APACrefauthors}%
\unskip\
\newblock
\APACrefYearMonthDay{2011}{}{}.
\newblock
{\BBOQ}\APACrefatitle {Bayesian inference for the spatial random effects model}
  {Bayesian inference for the spatial random effects model}.{\BBCQ}
\newblock
\APACjournalVolNumPages{Journal of the American Statistical
  Association}{106}{495}{972--983}.
\PrintBackRefs{\CurrentBib}

\bibitem [\protect \citeauthoryear {%
Kuiper%
}{%
Kuiper%
}{%
{\protect \APACyear {2020}}%
}]{%
kuiper2020}
\APACinsertmetastar {%
kuiper2020}%
\begin{APACrefauthors}%
Kuiper, R.%
\end{APACrefauthors}%
\unskip\
\newblock
\APACrefYearMonthDay{2020}{}{}.
\newblock
\APACrefbtitle {ChiBarSq.DiffTest: Chi-bar-square difference test of the
  RI-CLPM versus the CLPM and more general} {Chibarsq.difftest: Chi-bar-square
  difference test of the ri-clpm versus the clpm and more general}\
  \APACbVolEdTR{}{\BTR{}}.
\newblock
\APACaddressInstitution{r.m.kuiper@uu.nl}{Utrecht University}.
\PrintBackRefs{\CurrentBib}

\bibitem [\protect \citeauthoryear {%
Leamer%
}{%
Leamer%
}{%
{\protect \APACyear {2010}}%
}]{%
leamer2010tantalus}
\APACinsertmetastar {%
leamer2010tantalus}%
\begin{APACrefauthors}%
Leamer, E\BPBI E.%
\end{APACrefauthors}%
\unskip\
\newblock
\APACrefYearMonthDay{2010}{}{}.
\newblock
{\BBOQ}\APACrefatitle {Tantalus on the Road to {A}symptopia} {Tantalus on the
  road to {A}symptopia}.{\BBCQ}
\newblock
\APACjournalVolNumPages{Journal of Economic Perspectives}{24}{2}{31--46}.
\PrintBackRefs{\CurrentBib}

\bibitem [\protect \citeauthoryear {%
C\BPBI C.~Liu%
\ \BBA {} Aitkin%
}{%
C\BPBI C.~Liu%
\ \BBA {} Aitkin%
}{%
{\protect \APACyear {2008}}%
}]{%
liu2008bayes}
\APACinsertmetastar {%
liu2008bayes}%
\begin{APACrefauthors}%
Liu, C\BPBI C.%
\BCBT {}\ \BBA {} Aitkin, M.%
\end{APACrefauthors}%
\unskip\
\newblock
\APACrefYearMonthDay{2008}{}{}.
\newblock
{\BBOQ}\APACrefatitle {Bayes factors: Prior sensitivity and model
  generalizability} {Bayes factors: Prior sensitivity and model
  generalizability}.{\BBCQ}
\newblock
\APACjournalVolNumPages{Journal of Mathematical Psychology}{52}{6}{362--375}.
\PrintBackRefs{\CurrentBib}

\bibitem [\protect \citeauthoryear {%
S.~Liu%
, Kuppens%
\BCBL {}\ \BBA {} Bringmann%
}{%
S.~Liu%
\ \protect \BOthers {.}}{%
{\protect \APACyear {2019}}%
}]{%
liu2019use}
\APACinsertmetastar {%
liu2019use}%
\begin{APACrefauthors}%
Liu, S.%
, Kuppens, P.%
\BCBL {}\ \BBA {} Bringmann, L.%
\end{APACrefauthors}%
\unskip\
\newblock
\APACrefYearMonthDay{2019}{}{}.
\newblock
\APACrefbtitle {On the use of empirical Bayes estimates as measures of
  individual traits. Assessment. Advance online publication.} {On the use of
  empirical bayes estimates as measures of individual traits. assessment.
  advance online publication.}
\PrintBackRefs{\CurrentBib}

\bibitem [\protect \citeauthoryear {%
Ludbrook%
}{%
Ludbrook%
}{%
{\protect \APACyear {2011}}%
}]{%
ludbrook2011there}
\APACinsertmetastar {%
ludbrook2011there}%
\begin{APACrefauthors}%
Ludbrook, J.%
\end{APACrefauthors}%
\unskip\
\newblock
\APACrefYearMonthDay{2011}{}{}.
\newblock
{\BBOQ}\APACrefatitle {Is there still a place for Pearson's chi-squared test
  and Fisher's exact test in surgical research?} {Is there still a place for
  pearson's chi-squared test and fisher's exact test in surgical
  research?}{\BBCQ}
\newblock
\APACjournalVolNumPages{ANZ journal of surgery}{81}{12}{923--926}.
\PrintBackRefs{\CurrentBib}

\bibitem [\protect \citeauthoryear {%
Mackinnon%
, Curtis%
\BCBL {}\ \BBA {} O'Connor%
}{%
Mackinnon%
\ \protect \BOthers {.}}{%
{\protect \APACyear {2022}}%
}]{%
mackinnon2022tutorial}
\APACinsertmetastar {%
mackinnon2022tutorial}%
\begin{APACrefauthors}%
Mackinnon, S.%
, Curtis, R.%
\BCBL {}\ \BBA {} O'Connor, R.%
\end{APACrefauthors}%
\unskip\
\newblock
\APACrefYearMonthDay{2022}{}{}.
\newblock
{\BBOQ}\APACrefatitle {A tutorial in longitudinal measurement invariance and
  cross-lagged panel models using lavaan} {A tutorial in longitudinal
  measurement invariance and cross-lagged panel models using lavaan}.{\BBCQ}
\newblock
\APACjournalVolNumPages{Meta-Psychology}{6}{}{}.
\PrintBackRefs{\CurrentBib}

\bibitem [\protect \citeauthoryear {%
Marin%
\ \BBA {} Robert%
}{%
Marin%
\ \BBA {} Robert%
}{%
{\protect \APACyear {2010}}%
}]{%
marin2010resolving}
\APACinsertmetastar {%
marin2010resolving}%
\begin{APACrefauthors}%
Marin, J\BHBI M.%
\BCBT {}\ \BBA {} Robert, C\BPBI P.%
\end{APACrefauthors}%
\unskip\
\newblock
\APACrefYearMonthDay{2010}{}{}.
\newblock
{\BBOQ}\APACrefatitle {On resolving the Savage--Dickey paradox} {On resolving
  the savage--dickey paradox}.{\BBCQ}
\newblock
\APACjournalVolNumPages{Electronic Journal of Statistics}{4}{}{643--654}.
\PrintBackRefs{\CurrentBib}

\bibitem [\protect \citeauthoryear {%
McCullagh%
}{%
McCullagh%
}{%
{\protect \APACyear {2019}}%
}]{%
mccullagh2019generalized}
\APACinsertmetastar {%
mccullagh2019generalized}%
\begin{APACrefauthors}%
McCullagh, P.%
\end{APACrefauthors}%
\unskip\
\newblock
\APACrefYear{2019}.
\newblock
\APACrefbtitle {Generalized linear models} {Generalized linear models}.
\newblock
\APACaddressPublisher{}{Routledge}.
\PrintBackRefs{\CurrentBib}

\bibitem [\protect \citeauthoryear {%
McNeish%
, Mackinnon%
, Marsch%
\BCBL {}\ \BBA {} Poldrack%
}{%
McNeish%
\ \protect \BOthers {.}}{%
{\protect \APACyear {2021}}%
}]{%
mcneish2021measurement}
\APACinsertmetastar {%
mcneish2021measurement}%
\begin{APACrefauthors}%
McNeish, D.%
, Mackinnon, D\BPBI P.%
, Marsch, L\BPBI A.%
\BCBL {}\ \BBA {} Poldrack, R\BPBI A.%
\end{APACrefauthors}%
\unskip\
\newblock
\APACrefYearMonthDay{2021}{}{}.
\newblock
{\BBOQ}\APACrefatitle {Measurement in intensive longitudinal data} {Measurement
  in intensive longitudinal data}.{\BBCQ}
\newblock
\APACjournalVolNumPages{Structural equation modeling: a multidisciplinary
  journal}{28}{5}{807--822}.
\PrintBackRefs{\CurrentBib}

\bibitem [\protect \citeauthoryear {%
Menegaki%
}{%
Menegaki%
}{%
{\protect \APACyear {2011}}%
}]{%
menegaki2011growth}
\APACinsertmetastar {%
menegaki2011growth}%
\begin{APACrefauthors}%
Menegaki, A\BPBI N.%
\end{APACrefauthors}%
\unskip\
\newblock
\APACrefYearMonthDay{2011}{}{}.
\newblock
{\BBOQ}\APACrefatitle {Growth and renewable energy in Europe: A random effect
  model with evidence for neutrality hypothesis} {Growth and renewable energy
  in europe: A random effect model with evidence for neutrality
  hypothesis}.{\BBCQ}
\newblock
\APACjournalVolNumPages{Energy economics}{33}{2}{257--263}.
\PrintBackRefs{\CurrentBib}

\bibitem [\protect \citeauthoryear {%
Merkle%
, Fitzsimmons%
, Uanhoro%
\BCBL {}\ \BBA {} Goodrich%
}{%
Merkle%
\ \protect \BOthers {.}}{%
{\protect \APACyear {2020}}%
}]{%
merkle2020efficient}
\APACinsertmetastar {%
merkle2020efficient}%
\begin{APACrefauthors}%
Merkle, E\BPBI C.%
, Fitzsimmons, E.%
, Uanhoro, J.%
\BCBL {}\ \BBA {} Goodrich, B.%
\end{APACrefauthors}%
\unskip\
\newblock
\APACrefYearMonthDay{2020}{}{}.
\newblock
{\BBOQ}\APACrefatitle {Efficient Bayesian structural equation modeling in Stan}
  {Efficient bayesian structural equation modeling in stan}.{\BBCQ}
\newblock
\APACjournalVolNumPages{arXiv preprint arXiv:2008.07733}{}{}{}.
\PrintBackRefs{\CurrentBib}

\bibitem [\protect \citeauthoryear {%
Merkle%
\ \BBA {} Rosseel%
}{%
Merkle%
\ \BBA {} Rosseel%
}{%
{\protect \APACyear {2015}}%
}]{%
merkle2015blavaan}
\APACinsertmetastar {%
merkle2015blavaan}%
\begin{APACrefauthors}%
Merkle, E\BPBI C.%
\BCBT {}\ \BBA {} Rosseel, Y.%
\end{APACrefauthors}%
\unskip\
\newblock
\APACrefYearMonthDay{2015}{}{}.
\newblock
{\BBOQ}\APACrefatitle {blavaan: Bayesian structural equation models via
  parameter expansion} {blavaan: Bayesian structural equation models via
  parameter expansion}.{\BBCQ}
\newblock
\APACjournalVolNumPages{arXiv preprint arXiv:1511.05604}{}{}{}.
\PrintBackRefs{\CurrentBib}

\bibitem [\protect \citeauthoryear {%
J.~Mulder%
}{%
J.~Mulder%
}{%
{\protect \APACyear {2023}}%
}]{%
mulder2023bayesian}
\APACinsertmetastar {%
mulder2023bayesian}%
\begin{APACrefauthors}%
Mulder, J.%
\end{APACrefauthors}%
\unskip\
\newblock
\APACrefYearMonthDay{2023}{}{}.
\newblock
{\BBOQ}\APACrefatitle {Bayesian testing of linear versus nonlinear effects
  using Gaussian process priors} {Bayesian testing of linear versus nonlinear
  effects using gaussian process priors}.{\BBCQ}
\newblock
\APACjournalVolNumPages{The American Statistician}{77}{1}{1--11}.
\PrintBackRefs{\CurrentBib}

\bibitem [\protect \citeauthoryear {%
J.~Mulder%
\ \BBA {} Fox%
}{%
J.~Mulder%
\ \BBA {} Fox%
}{%
{\protect \APACyear {2013}}%
}]{%
mulder2013bayesian}
\APACinsertmetastar {%
mulder2013bayesian}%
\begin{APACrefauthors}%
Mulder, J.%
\BCBT {}\ \BBA {} Fox, J\BHBI P.%
\end{APACrefauthors}%
\unskip\
\newblock
\APACrefYearMonthDay{2013}{}{}.
\newblock
{\BBOQ}\APACrefatitle {Bayesian tests on components of the compound symmetry
  covariance matrix} {Bayesian tests on components of the compound symmetry
  covariance matrix}.{\BBCQ}
\newblock
\APACjournalVolNumPages{Statistics and Computing}{23}{1}{109--122}.
\PrintBackRefs{\CurrentBib}

\bibitem [\protect \citeauthoryear {%
J.~Mulder%
\ \BBA {} Fox%
}{%
J.~Mulder%
\ \BBA {} Fox%
}{%
{\protect \APACyear {2019}}%
}]{%
mulder2019bayes}
\APACinsertmetastar {%
mulder2019bayes}%
\begin{APACrefauthors}%
Mulder, J.%
\BCBT {}\ \BBA {} Fox, J\BHBI P.%
\end{APACrefauthors}%
\unskip\
\newblock
\APACrefYearMonthDay{2019}{}{}.
\newblock
{\BBOQ}\APACrefatitle {Bayes factor testing of multiple intraclass
  correlations} {Bayes factor testing of multiple intraclass
  correlations}.{\BBCQ}
\newblock
\APACjournalVolNumPages{Bayesian Analysis}{}{}{}.
\PrintBackRefs{\CurrentBib}

\bibitem [\protect \citeauthoryear {%
J.~Mulder%
, Wagenmakers%
\BCBL {}\ \BBA {} Marsman%
}{%
J.~Mulder%
\ \protect \BOthers {.}}{%
{\protect \APACyear {2022}}%
}]{%
mulder2022generalization}
\APACinsertmetastar {%
mulder2022generalization}%
\begin{APACrefauthors}%
Mulder, J.%
, Wagenmakers, E\BHBI J.%
\BCBL {}\ \BBA {} Marsman, M.%
\end{APACrefauthors}%
\unskip\
\newblock
\APACrefYearMonthDay{2022}{}{}.
\newblock
{\BBOQ}\APACrefatitle {A generalization of the {S}avage-{D}ickey density ratio
  for testing equality and order constrained hypotheses} {A generalization of
  the {S}avage-{D}ickey density ratio for testing equality and order
  constrained hypotheses}.{\BBCQ}
\newblock
\APACjournalVolNumPages{The American Statistician}{76}{2}{102--109}.
\PrintBackRefs{\CurrentBib}

\bibitem [\protect \citeauthoryear {%
J\BPBI D.~Mulder%
\ \BBA {} Hamaker%
}{%
J\BPBI D.~Mulder%
\ \BBA {} Hamaker%
}{%
{\protect \APACyear {2021}}%
}]{%
mulder2021three}
\APACinsertmetastar {%
mulder2021three}%
\begin{APACrefauthors}%
Mulder, J\BPBI D.%
\BCBT {}\ \BBA {} Hamaker, E\BPBI L.%
\end{APACrefauthors}%
\unskip\
\newblock
\APACrefYearMonthDay{2021}{}{}.
\newblock
{\BBOQ}\APACrefatitle {Three extensions of the random intercept cross-lagged
  panel model} {Three extensions of the random intercept cross-lagged panel
  model}.{\BBCQ}
\newblock
\APACjournalVolNumPages{Structural Equation Modeling: A Multidisciplinary
  Journal}{28}{4}{638--648}.
\PrintBackRefs{\CurrentBib}

\bibitem [\protect \citeauthoryear {%
Nielsen%
, Smink%
\BCBL {}\ \BBA {} Fox%
}{%
Nielsen%
\ \protect \BOthers {.}}{%
{\protect \APACyear {2021}}%
}]{%
nielsen2021small}
\APACinsertmetastar {%
nielsen2021small}%
\begin{APACrefauthors}%
Nielsen, N\BPBI M.%
, Smink, W\BPBI A.%
\BCBL {}\ \BBA {} Fox, J\BHBI P.%
\end{APACrefauthors}%
\unskip\
\newblock
\APACrefYearMonthDay{2021}{}{}.
\newblock
{\BBOQ}\APACrefatitle {Small and negative correlations among clustered
  observations: Limitations of the linear mixed effects model} {Small and
  negative correlations among clustered observations: Limitations of the linear
  mixed effects model}.{\BBCQ}
\newblock
\APACjournalVolNumPages{Behaviormetrika}{48}{}{51--77}.
\PrintBackRefs{\CurrentBib}

\bibitem [\protect \citeauthoryear {%
O'Hagan%
}{%
O'Hagan%
}{%
{\protect \APACyear {1995}}%
}]{%
o1995fractional}
\APACinsertmetastar {%
o1995fractional}%
\begin{APACrefauthors}%
O'Hagan, A.%
\end{APACrefauthors}%
\unskip\
\newblock
\APACrefYearMonthDay{1995}{}{}.
\newblock
{\BBOQ}\APACrefatitle {Fractional Bayes factors for model comparison}
  {Fractional bayes factors for model comparison}.{\BBCQ}
\newblock
\APACjournalVolNumPages{Journal of the Royal Statistical Society: Series B
  (Methodological)}{57}{1}{99--118}.
\PrintBackRefs{\CurrentBib}

\bibitem [\protect \citeauthoryear {%
Pauler%
, Wakefield%
\BCBL {}\ \BBA {} Kass%
}{%
Pauler%
\ \protect \BOthers {.}}{%
{\protect \APACyear {1999}}%
}]{%
pauler1999bayes}
\APACinsertmetastar {%
pauler1999bayes}%
\begin{APACrefauthors}%
Pauler, D\BPBI K.%
, Wakefield, J\BPBI C.%
\BCBL {}\ \BBA {} Kass, R\BPBI E.%
\end{APACrefauthors}%
\unskip\
\newblock
\APACrefYearMonthDay{1999}{}{}.
\newblock
{\BBOQ}\APACrefatitle {Bayes factors and approximations for variance component
  models} {Bayes factors and approximations for variance component
  models}.{\BBCQ}
\newblock
\APACjournalVolNumPages{Journal of the American Statistical
  Association}{94}{448}{1242--1253}.
\PrintBackRefs{\CurrentBib}

\bibitem [\protect \citeauthoryear {%
Plummer%
}{%
Plummer%
}{%
{\protect \APACyear {2012}}%
}]{%
plummer2012jags}
\APACinsertmetastar {%
plummer2012jags}%
\begin{APACrefauthors}%
Plummer, M.%
\end{APACrefauthors}%
\unskip\
\newblock
\APACrefYearMonthDay{2012}{}{}.
\newblock
\APACrefbtitle {JAGS Version 3.3. 0 user manual.} {Jags version 3.3. 0 user
  manual.}
\newblock
\APACaddressPublisher{}{Lyon, France}.
\PrintBackRefs{\CurrentBib}

\bibitem [\protect \citeauthoryear {%
Raftery%
}{%
Raftery%
}{%
{\protect \APACyear {1995}}%
}]{%
raftery1995bayesian}
\APACinsertmetastar {%
raftery1995bayesian}%
\begin{APACrefauthors}%
Raftery, A\BPBI E.%
\end{APACrefauthors}%
\unskip\
\newblock
\APACrefYearMonthDay{1995}{}{}.
\newblock
{\BBOQ}\APACrefatitle {Bayesian model selection in social research} {Bayesian
  model selection in social research}.{\BBCQ}
\newblock
\APACjournalVolNumPages{Sociological methodology}{}{}{111--163}.
\PrintBackRefs{\CurrentBib}

\bibitem [\protect \citeauthoryear {%
Roback%
\ \BBA {} Legler%
}{%
Roback%
\ \BBA {} Legler%
}{%
{\protect \APACyear {2021}}%
}]{%
roback2021beyond}
\APACinsertmetastar {%
roback2021beyond}%
\begin{APACrefauthors}%
Roback, P.%
\BCBT {}\ \BBA {} Legler, J.%
\end{APACrefauthors}%
\unskip\
\newblock
\APACrefYear{2021}.
\newblock
\APACrefbtitle {Beyond multiple linear regression: applied generalized linear
  models and multilevel models in R} {Beyond multiple linear regression:
  applied generalized linear models and multilevel models in r}.
\newblock
\APACaddressPublisher{}{Chapman and Hall/CRC}.
\PrintBackRefs{\CurrentBib}

\bibitem [\protect \citeauthoryear {%
Rosseel%
}{%
Rosseel%
}{%
{\protect \APACyear {2012}}%
}]{%
rosseel2012lavaan}
\APACinsertmetastar {%
rosseel2012lavaan}%
\begin{APACrefauthors}%
Rosseel, Y.%
\end{APACrefauthors}%
\unskip\
\newblock
\APACrefYearMonthDay{2012}{}{}.
\newblock
{\BBOQ}\APACrefatitle {lavaan: An R package for structural equation modeling}
  {lavaan: An r package for structural equation modeling}.{\BBCQ}
\newblock
\APACjournalVolNumPages{Journal of statistical software}{48}{}{1--36}.
\PrintBackRefs{\CurrentBib}

\bibitem [\protect \citeauthoryear {%
Schaffer%
}{%
Schaffer%
}{%
{\protect \APACyear {1993}}%
}]{%
schaffer1993selecting}
\APACinsertmetastar {%
schaffer1993selecting}%
\begin{APACrefauthors}%
Schaffer, C.%
\end{APACrefauthors}%
\unskip\
\newblock
\APACrefYearMonthDay{1993}{}{}.
\newblock
{\BBOQ}\APACrefatitle {Selecting a classification method by cross-validation}
  {Selecting a classification method by cross-validation}.{\BBCQ}
\newblock
\APACjournalVolNumPages{Machine learning}{13}{}{135--143}.
\PrintBackRefs{\CurrentBib}

\bibitem [\protect \citeauthoryear {%
Schwarz%
}{%
Schwarz%
}{%
{\protect \APACyear {1978}}%
}]{%
schwarz1978estimating}
\APACinsertmetastar {%
schwarz1978estimating}%
\begin{APACrefauthors}%
Schwarz, G.%
\end{APACrefauthors}%
\unskip\
\newblock
\APACrefYearMonthDay{1978}{}{}.
\newblock
{\BBOQ}\APACrefatitle {Estimating the dimension of a model} {Estimating the
  dimension of a model}.{\BBCQ}
\newblock
\APACjournalVolNumPages{The annals of statistics}{}{}{461--464}.
\PrintBackRefs{\CurrentBib}

\bibitem [\protect \citeauthoryear {%
Sellke%
, Bayarri%
\BCBL {}\ \BBA {} Berger%
}{%
Sellke%
\ \protect \BOthers {.}}{%
{\protect \APACyear {2001}}%
}]{%
sellke2001calibration}
\APACinsertmetastar {%
sellke2001calibration}%
\begin{APACrefauthors}%
Sellke, T.%
, Bayarri, M\BPBI J.%
\BCBL {}\ \BBA {} Berger, J\BPBI O.%
\end{APACrefauthors}%
\unskip\
\newblock
\APACrefYearMonthDay{2001}{}{}.
\newblock
{\BBOQ}\APACrefatitle {Calibration of $\rho$ values for testing precise null
  hypotheses} {Calibration of $\rho$ values for testing precise null
  hypotheses}.{\BBCQ}
\newblock
\APACjournalVolNumPages{The American Statistician}{55}{1}{62--71}.
\PrintBackRefs{\CurrentBib}

\bibitem [\protect \citeauthoryear {%
Shao%
}{%
Shao%
}{%
{\protect \APACyear {1993}}%
}]{%
shao1993linear}
\APACinsertmetastar {%
shao1993linear}%
\begin{APACrefauthors}%
Shao, J.%
\end{APACrefauthors}%
\unskip\
\newblock
\APACrefYearMonthDay{1993}{}{}.
\newblock
{\BBOQ}\APACrefatitle {Linear model selection by cross-validation} {Linear
  model selection by cross-validation}.{\BBCQ}
\newblock
\APACjournalVolNumPages{Journal of the American statistical
  Association}{88}{422}{486--494}.
\PrintBackRefs{\CurrentBib}

\bibitem [\protect \citeauthoryear {%
Sharpe%
}{%
Sharpe%
}{%
{\protect \APACyear {2015}}%
}]{%
sharpe2015chi}
\APACinsertmetastar {%
sharpe2015chi}%
\begin{APACrefauthors}%
Sharpe, D.%
\end{APACrefauthors}%
\unskip\
\newblock
\APACrefYearMonthDay{2015}{}{}.
\newblock
{\BBOQ}\APACrefatitle {Chi-square test is statistically significant: Now what?}
  {Chi-square test is statistically significant: Now what?}{\BBCQ}
\newblock
\APACjournalVolNumPages{Practical Assessment, Research, and
  Evaluation}{20}{1}{8}.
\PrintBackRefs{\CurrentBib}

\bibitem [\protect \citeauthoryear {%
Sinharay%
\ \BBA {} Stern%
}{%
Sinharay%
\ \BBA {} Stern%
}{%
{\protect \APACyear {2002}}%
}]{%
sinharay2002sensitivity}
\APACinsertmetastar {%
sinharay2002sensitivity}%
\begin{APACrefauthors}%
Sinharay, S.%
\BCBT {}\ \BBA {} Stern, H\BPBI S.%
\end{APACrefauthors}%
\unskip\
\newblock
\APACrefYearMonthDay{2002}{}{}.
\newblock
{\BBOQ}\APACrefatitle {On the sensitivity of Bayes factors to the prior
  distributions} {On the sensitivity of bayes factors to the prior
  distributions}.{\BBCQ}
\newblock
\APACjournalVolNumPages{The American Statistician}{56}{3}{196--201}.
\PrintBackRefs{\CurrentBib}

\bibitem [\protect \citeauthoryear {%
Spiegelhalter%
, Best%
, Carlin%
\BCBL {}\ \BBA {} Van Der~Linde%
}{%
Spiegelhalter%
\ \protect \BOthers {.}}{%
{\protect \APACyear {2002}}%
}]{%
spiegelhalter2002bayesian}
\APACinsertmetastar {%
spiegelhalter2002bayesian}%
\begin{APACrefauthors}%
Spiegelhalter, D\BPBI J.%
, Best, N\BPBI G.%
, Carlin, B\BPBI P.%
\BCBL {}\ \BBA {} Van Der~Linde, A.%
\end{APACrefauthors}%
\unskip\
\newblock
\APACrefYearMonthDay{2002}{}{}.
\newblock
{\BBOQ}\APACrefatitle {Bayesian measures of model complexity and fit} {Bayesian
  measures of model complexity and fit}.{\BBCQ}
\newblock
\APACjournalVolNumPages{Journal of the royal statistical society: Series b
  (statistical methodology)}{64}{4}{583--639}.
\PrintBackRefs{\CurrentBib}

\bibitem [\protect \citeauthoryear {%
{Stan Development Team}%
}{%
{Stan Development Team}%
}{%
{\protect \APACyear {2024}}%
}]{%
rstan2024}
\APACinsertmetastar {%
rstan2024}%
\begin{APACrefauthors}%
{Stan Development Team}.%
\end{APACrefauthors}%
\unskip\
\newblock
\APACrefYearMonthDay{2024}{}{}.
\newblock
\APACrefbtitle {{RStan}: the {R} interface to {Stan}.} {{RStan}: the {R}
  interface to {Stan}.}
\newblock
\begin{APACrefURL} \url{https://mc-stan.org/} \end{APACrefURL}
\newblock
\APACrefnote{R package version 2.32.5}
\PrintBackRefs{\CurrentBib}

\bibitem [\protect \citeauthoryear {%
Stoel%
, Garre%
, Dolan%
\BCBL {}\ \BBA {} Van Den~Wittenboer%
}{%
Stoel%
\ \protect \BOthers {.}}{%
{\protect \APACyear {2006}}%
}]{%
stoel2006likelihood}
\APACinsertmetastar {%
stoel2006likelihood}%
\begin{APACrefauthors}%
Stoel, R\BPBI D.%
, Garre, F\BPBI G.%
, Dolan, C.%
\BCBL {}\ \BBA {} Van Den~Wittenboer, G.%
\end{APACrefauthors}%
\unskip\
\newblock
\APACrefYearMonthDay{2006}{}{}.
\newblock
{\BBOQ}\APACrefatitle {On the likelihood ratio test in structural equation
  modeling when parameters are subject to boundary constraints.} {On the
  likelihood ratio test in structural equation modeling when parameters are
  subject to boundary constraints.}{\BBCQ}
\newblock
\APACjournalVolNumPages{Psychological methods}{11}{4}{439}.
\PrintBackRefs{\CurrentBib}

\bibitem [\protect \citeauthoryear {%
Stram%
\ \BBA {} Lee%
}{%
Stram%
\ \BBA {} Lee%
}{%
{\protect \APACyear {1994}}%
}]{%
stram1994variance}
\APACinsertmetastar {%
stram1994variance}%
\begin{APACrefauthors}%
Stram, D\BPBI O.%
\BCBT {}\ \BBA {} Lee, J\BPBI W.%
\end{APACrefauthors}%
\unskip\
\newblock
\APACrefYearMonthDay{1994}{}{}.
\newblock
{\BBOQ}\APACrefatitle {Variance components testing in the longitudinal mixed
  effects model} {Variance components testing in the longitudinal mixed effects
  model}.{\BBCQ}
\newblock
\APACjournalVolNumPages{Biometrics}{}{}{1171--1177}.
\PrintBackRefs{\CurrentBib}

\bibitem [\protect \citeauthoryear {%
Ten~Have%
\ \BBA {} Localio%
}{%
Ten~Have%
\ \BBA {} Localio%
}{%
{\protect \APACyear {1999}}%
}]{%
ten1999empirical}
\APACinsertmetastar {%
ten1999empirical}%
\begin{APACrefauthors}%
Ten~Have, T\BPBI R.%
\BCBT {}\ \BBA {} Localio, A\BPBI R.%
\end{APACrefauthors}%
\unskip\
\newblock
\APACrefYearMonthDay{1999}{}{}.
\newblock
{\BBOQ}\APACrefatitle {Empirical Bayes estimation of random effects parameters
  in mixed effects logistic regression models} {Empirical bayes estimation of
  random effects parameters in mixed effects logistic regression
  models}.{\BBCQ}
\newblock
\APACjournalVolNumPages{Biometrics}{55}{4}{1022--1029}.
\PrintBackRefs{\CurrentBib}

\bibitem [\protect \citeauthoryear {%
Terrin%
, Schmid%
\BCBL {}\ \BBA {} Lau%
}{%
Terrin%
\ \protect \BOthers {.}}{%
{\protect \APACyear {2005}}%
}]{%
terrin2005empirical}
\APACinsertmetastar {%
terrin2005empirical}%
\begin{APACrefauthors}%
Terrin, N.%
, Schmid, C\BPBI H.%
\BCBL {}\ \BBA {} Lau, J.%
\end{APACrefauthors}%
\unskip\
\newblock
\APACrefYearMonthDay{2005}{}{}.
\newblock
{\BBOQ}\APACrefatitle {In an empirical evaluation of the funnel plot,
  researchers could not visually identify publication bias} {In an empirical
  evaluation of the funnel plot, researchers could not visually identify
  publication bias}.{\BBCQ}
\newblock
\APACjournalVolNumPages{Journal of clinical epidemiology}{58}{9}{894--901}.
\PrintBackRefs{\CurrentBib}

\bibitem [\protect \citeauthoryear {%
Tukey%
}{%
Tukey%
}{%
{\protect \APACyear {1977}}%
}]{%
tukey1977exploratory}
\APACinsertmetastar {%
tukey1977exploratory}%
\begin{APACrefauthors}%
Tukey, J\BPBI W.%
\end{APACrefauthors}%
\unskip\
\newblock
\APACrefYear{1977}.
\newblock
\APACrefbtitle {Exploratory data analysis} {Exploratory data analysis}\
  (\BVOL~2).
\newblock
\APACaddressPublisher{}{Reading, MA}.
\PrintBackRefs{\CurrentBib}

\bibitem [\protect \citeauthoryear {%
Van~Erp%
, Oberski%
\BCBL {}\ \BBA {} Mulder%
}{%
Van~Erp%
\ \protect \BOthers {.}}{%
{\protect \APACyear {2019}}%
}]{%
van2019shrinkage}
\APACinsertmetastar {%
van2019shrinkage}%
\begin{APACrefauthors}%
Van~Erp, S.%
, Oberski, D\BPBI L.%
\BCBL {}\ \BBA {} Mulder, J.%
\end{APACrefauthors}%
\unskip\
\newblock
\APACrefYearMonthDay{2019}{}{}.
\newblock
{\BBOQ}\APACrefatitle {Shrinkage priors for Bayesian penalized regression}
  {Shrinkage priors for bayesian penalized regression}.{\BBCQ}
\newblock
\APACjournalVolNumPages{Journal of Mathematical Psychology}{89}{}{31--50}.
\PrintBackRefs{\CurrentBib}

\bibitem [\protect \citeauthoryear {%
Vanpaemel%
}{%
Vanpaemel%
}{%
{\protect \APACyear {2010}}%
}]{%
vanpaemel2010prior}
\APACinsertmetastar {%
vanpaemel2010prior}%
\begin{APACrefauthors}%
Vanpaemel, W.%
\end{APACrefauthors}%
\unskip\
\newblock
\APACrefYearMonthDay{2010}{}{}.
\newblock
{\BBOQ}\APACrefatitle {Prior sensitivity in theory testing: An apologia for the
  Bayes factor} {Prior sensitivity in theory testing: An apologia for the bayes
  factor}.{\BBCQ}
\newblock
\APACjournalVolNumPages{Journal of Mathematical Psychology}{54}{6}{491--498}.
\PrintBackRefs{\CurrentBib}

\bibitem [\protect \citeauthoryear {%
Vehtari%
\ \protect \BOthers {.}}{%
Vehtari%
\ \protect \BOthers {.}}{%
{\protect \APACyear {2020}}%
}]{%
vehtari2020loo}
\APACinsertmetastar {%
vehtari2020loo}%
\begin{APACrefauthors}%
Vehtari, A.%
, Gabry, J.%
, Magnusson, M.%
, Yao, Y.%
, B{\"u}rkner, P\BHBI C.%
, Paananen, T.%
\BCBL {}\ \BBA {} Gelman, A.%
\end{APACrefauthors}%
\unskip\
\newblock
\APACrefYearMonthDay{2020}{}{}.
\newblock
{\BBOQ}\APACrefatitle {loo: Efficient leave-one-out cross-validation and WAIC
  for Bayesian models} {loo: Efficient leave-one-out cross-validation and waic
  for bayesian models}.{\BBCQ}
\newblock
\APACjournalVolNumPages{R package version}{2}{1}{12}.
\PrintBackRefs{\CurrentBib}

\bibitem [\protect \citeauthoryear {%
Vehtari%
\ \BBA {} Ojanen%
}{%
Vehtari%
\ \BBA {} Ojanen%
}{%
{\protect \APACyear {2012}}%
}]{%
vehtari2012survey}
\APACinsertmetastar {%
vehtari2012survey}%
\begin{APACrefauthors}%
Vehtari, A.%
\BCBT {}\ \BBA {} Ojanen, J.%
\end{APACrefauthors}%
\unskip\
\newblock
\APACrefYearMonthDay{2012}{}{}.
\newblock
{\BBOQ}\APACrefatitle {A survey of Bayesian predictive methods for model
  assessment, selection and comparison} {A survey of bayesian predictive
  methods for model assessment, selection and comparison}.{\BBCQ}
\newblock
\APACjournalVolNumPages{Statistics and Computing}{}{}{}.
\PrintBackRefs{\CurrentBib}

\bibitem [\protect \citeauthoryear {%
Verbeke%
\ \BBA {} Molenberghs%
}{%
Verbeke%
\ \BBA {} Molenberghs%
}{%
{\protect \APACyear {2012}}%
}]{%
verbeke2012linear}
\APACinsertmetastar {%
verbeke2012linear}%
\begin{APACrefauthors}%
Verbeke, G.%
\BCBT {}\ \BBA {} Molenberghs, G.%
\end{APACrefauthors}%
\unskip\
\newblock
\APACrefYear{2012}.
\newblock
\APACrefbtitle {Linear mixed models in practice: a SAS-oriented approach}
  {Linear mixed models in practice: a sas-oriented approach}\ (\BVOL~126).
\newblock
\APACaddressPublisher{}{Springer Science \& Business Media}.
\PrintBackRefs{\CurrentBib}

\bibitem [\protect \citeauthoryear {%
Verdinelli%
\ \BBA {} Wasserman%
}{%
Verdinelli%
\ \BBA {} Wasserman%
}{%
{\protect \APACyear {1995}}%
}]{%
verdinelli1995computing}
\APACinsertmetastar {%
verdinelli1995computing}%
\begin{APACrefauthors}%
Verdinelli, I.%
\BCBT {}\ \BBA {} Wasserman, L.%
\end{APACrefauthors}%
\unskip\
\newblock
\APACrefYearMonthDay{1995}{}{}.
\newblock
{\BBOQ}\APACrefatitle {Computing Bayes factors using a generalization of the
  Savage-Dickey density ratio} {Computing bayes factors using a generalization
  of the savage-dickey density ratio}.{\BBCQ}
\newblock
\APACjournalVolNumPages{Journal of the American Statistical
  Association}{90}{430}{614--618}.
\PrintBackRefs{\CurrentBib}

\bibitem [\protect \citeauthoryear {%
Verhagen%
\ \BBA {} Fox%
}{%
Verhagen%
\ \BBA {} Fox%
}{%
{\protect \APACyear {2013}}%
}]{%
verhagen2013bayesian}
\APACinsertmetastar {%
verhagen2013bayesian}%
\begin{APACrefauthors}%
Verhagen, A.%
\BCBT {}\ \BBA {} Fox, J.%
\end{APACrefauthors}%
\unskip\
\newblock
\APACrefYearMonthDay{2013}{}{}.
\newblock
{\BBOQ}\APACrefatitle {Bayesian tests of measurement invariance} {Bayesian
  tests of measurement invariance}.{\BBCQ}
\newblock
\APACjournalVolNumPages{British Journal of Mathematical and Statistical
  Psychology}{66}{3}{383--401}.
\PrintBackRefs{\CurrentBib}

\bibitem [\protect \citeauthoryear {%
Vieira%
, Leenders%
, McFarland%
\BCBL {}\ \BBA {} Mulder%
}{%
Vieira%
\ \protect \BOthers {.}}{%
{\protect \APACyear {2023}}%
}]{%
vieira2023bayesian}
\APACinsertmetastar {%
vieira2023bayesian}%
\begin{APACrefauthors}%
Vieira, F.%
, Leenders, R.%
, McFarland, D.%
\BCBL {}\ \BBA {} Mulder, J.%
\end{APACrefauthors}%
\unskip\
\newblock
\APACrefYearMonthDay{2023}{}{}.
\newblock
\APACrefbtitle {A {B}ayesian actor-oriented multilevel relational event model
  with hypothesis testing procedures.} {A {B}ayesian actor-oriented multilevel
  relational event model with hypothesis testing procedures.}
\PrintBackRefs{\CurrentBib}

\bibitem [\protect \citeauthoryear {%
Wagenmakers%
}{%
Wagenmakers%
}{%
{\protect \APACyear {2007}}%
}]{%
wagenmakers2007practical}
\APACinsertmetastar {%
wagenmakers2007practical}%
\begin{APACrefauthors}%
Wagenmakers, E\BHBI J.%
\end{APACrefauthors}%
\unskip\
\newblock
\APACrefYearMonthDay{2007}{}{}.
\newblock
{\BBOQ}\APACrefatitle {A practical solution to the pervasive problems of p
  values} {A practical solution to the pervasive problems of p values}.{\BBCQ}
\newblock
\APACjournalVolNumPages{Psychonomic bulletin \& review}{14}{5}{779--804}.
\PrintBackRefs{\CurrentBib}

\bibitem [\protect \citeauthoryear {%
Wagenmakers%
, Lodewyckx%
, Kuriyal%
\BCBL {}\ \BBA {} Grasman%
}{%
Wagenmakers%
\ \protect \BOthers {.}}{%
{\protect \APACyear {2010}}%
}]{%
wagenmakers2010bayesian}
\APACinsertmetastar {%
wagenmakers2010bayesian}%
\begin{APACrefauthors}%
Wagenmakers, E\BHBI J.%
, Lodewyckx, T.%
, Kuriyal, H.%
\BCBL {}\ \BBA {} Grasman, R.%
\end{APACrefauthors}%
\unskip\
\newblock
\APACrefYearMonthDay{2010}{}{}.
\newblock
{\BBOQ}\APACrefatitle {Bayesian hypothesis testing for psychologists: A
  tutorial on the Savage--Dickey method} {Bayesian hypothesis testing for
  psychologists: A tutorial on the savage--dickey method}.{\BBCQ}
\newblock
\APACjournalVolNumPages{Cognitive psychology}{60}{3}{158--189}.
\PrintBackRefs{\CurrentBib}

\bibitem [\protect \citeauthoryear {%
Wetzels%
, Grasman%
\BCBL {}\ \BBA {} Wagenmakers%
}{%
Wetzels%
\ \protect \BOthers {.}}{%
{\protect \APACyear {2010}}%
}]{%
wetzels2010encompassing}
\APACinsertmetastar {%
wetzels2010encompassing}%
\begin{APACrefauthors}%
Wetzels, R.%
, Grasman, R\BPBI P.%
\BCBL {}\ \BBA {} Wagenmakers, E\BHBI J.%
\end{APACrefauthors}%
\unskip\
\newblock
\APACrefYearMonthDay{2010}{}{}.
\newblock
{\BBOQ}\APACrefatitle {An encompassing prior generalization of the
  Savage--Dickey density ratio} {An encompassing prior generalization of the
  savage--dickey density ratio}.{\BBCQ}
\newblock
\APACjournalVolNumPages{Computational Statistics \& Data
  Analysis}{54}{9}{2094--2102}.
\PrintBackRefs{\CurrentBib}

\bibitem [\protect \citeauthoryear {%
Williams%
\ \BBA {} Rasmussen%
}{%
Williams%
\ \BBA {} Rasmussen%
}{%
{\protect \APACyear {2006}}%
}]{%
williams2006gaussian}
\APACinsertmetastar {%
williams2006gaussian}%
\begin{APACrefauthors}%
Williams, C\BPBI K.%
\BCBT {}\ \BBA {} Rasmussen, C\BPBI E.%
\end{APACrefauthors}%
\unskip\
\newblock
\APACrefYear{2006}.
\newblock
\APACrefbtitle {Gaussian processes for machine learning} {Gaussian processes
  for machine learning}\ (\BVOL~2)\ (\BNUM~3).
\newblock
\APACaddressPublisher{}{MIT press Cambridge, MA}.
\PrintBackRefs{\CurrentBib}

\bibitem [\protect \citeauthoryear {%
Wood%
}{%
Wood%
}{%
{\protect \APACyear {2017}}%
}]{%
wood2017generalized}
\APACinsertmetastar {%
wood2017generalized}%
\begin{APACrefauthors}%
Wood, S\BPBI N.%
\end{APACrefauthors}%
\unskip\
\newblock
\APACrefYear{2017}.
\newblock
\APACrefbtitle {Generalized additive models: an introduction with R}
  {Generalized additive models: an introduction with r}.
\newblock
\APACaddressPublisher{}{CRC press}.
\PrintBackRefs{\CurrentBib}

\bibitem [\protect \citeauthoryear {%
Zhang%
\ \BBA {} Lin%
}{%
Zhang%
\ \BBA {} Lin%
}{%
{\protect \APACyear {2008}}%
}]{%
zhang2008variance}
\APACinsertmetastar {%
zhang2008variance}%
\begin{APACrefauthors}%
Zhang, D.%
\BCBT {}\ \BBA {} Lin, X.%
\end{APACrefauthors}%
\unskip\
\newblock
\APACrefYearMonthDay{2008}{}{}.
\newblock
{\BBOQ}\APACrefatitle {Variance component testing in generalized linear mixed
  models for longitudinal/clustered data and other related topics} {Variance
  component testing in generalized linear mixed models for
  longitudinal/clustered data and other related topics}.{\BBCQ}
\newblock
\APACjournalVolNumPages{Random effect and latent variable model
  selection}{}{}{19--36}.
\PrintBackRefs{\CurrentBib}

\end{thebibliography}

\end{document}